\documentclass[fleqn,usenatbib]{mnras}

\usepackage{newtxtext,newtxmath}
\usepackage{xcolor}

\usepackage[T1]{fontenc}

\DeclareRobustCommand{\VAN}[3]{#2}
\let\VANthebibliography\thebibliography
\def\thebibliography{\DeclareRobustCommand{\VAN}[3]{##3}\VANthebibliography}

\usepackage{graphicx}	
\usepackage{amsmath}	
\usepackage{anyfontsize}

\newcommand{\gaia}{\textit{Gaia}}
\newcommand{\Teff}{\mbox{$T_{\mathrm{eff}}$}}
\newcommand{\Msun}{\mbox{M$_\odot$}}

\newcommand{\logg}{\mbox{$\log g$}}

\title[White dwarfs within 13 pc in the UV]{White dwarfs within 13 pc: insights from ultraviolet spectroscopy}

\author[M. W. O'Brien et al.]{Mairi~W.~O'Brien$^{1}$\thanks{E-mail: mairi.obrien1@gmail.com},
Pier-Emmanuel~Tremblay$^{1}$, Boris~T.~G\"ansicke$^{1}$, Mark~A.~Hollands$^{1}$,  
\newauthor Detlev~Koester$^{2}$, Snehalata~Sahu$^{1}$, Antoine~B\'edard$^{1}$, Andrew~M.~Buchan$^{1}$, Tim~Cunningham$^{3}$, 
\newauthor John~H.~Debes$^{4}$, Jay~Farihi$^{5}$, J.~J.~Hermes$^{6}$ and Piotr~M.~Kowalski$^{7}$
\\
$^{1}$ Department of Physics, University of Warwick, Coventry CV4 7AL, UK \\
$^{2}$ Institut f\"ur Theoretische Physik und Astrophysik, University of Kiel, 24098 Kiel, Germany \\
$^{3}$ Center for Astrophysics, Harvard \& Smithsonian, 60 Garden St., Cambridge, MA 02138, USA \\ 
$^{4}$ ESA for AURA, The Space Telescope Science Institute, 3700 San Martin Dr., Baltimore, MD 21218, USA \\
$^{5}$ Department of Physics and Astronomy, University College London, London WC1E 6BT, UK \\
$^{6}$ Department of Astronomy \& Institute for Astrophysical Research, Boston University, Boston, MA 02215, USA \\
$^{7}$ Institute of Energy Technologies $-$ Theory and Computation of Energy Materials (IET-3), Forschungszentrum J\"ulich GmbH, 52425
J\"ulich, Germany \\
}

\date{Accepted 2026 September 01. Received 2026 September 01; in original form 2026 July 07}

\pubyear{2026}

\begin{document}
\label{firstpage}
\pagerange{\pageref{firstpage}--\pageref{lastpage}}
\maketitle

\begin{abstract}
We present a comprehensive multi-wavelength spectroscopic and photometric analysis of the 44 confirmed white dwarfs within 13\,pc of the Sun. Combining flux-calibrated ultraviolet spectroscopy from the \textit{Hubble Space Telescope} (STIS and COS) with ground-based optical spectroscopy, as well as photometry from \gaia, 2MASS, and WISE, we employ a combined fitting method to calculate atmospheric parameters. Each white dwarf is fitted with a bespoke model depending on its detailed atmospheric composition. Two strongly magnetic stars could not be fitted due to the complex splitting of their spectral lines. We find a systematic discrepancy in hydrogen-atmosphere white dwarfs with effective temperatures below 10\,000\,K, where fits incorporating ultraviolet spectra result in effective temperatures that are 1\,--\,5\,per\,cent higher than those derived from optical and infrared photometry alone. We re-classify three helium-atmosphere white dwarfs as metal enriched following a magnesium detection in their near-ultraviolet spectra: WD\,0435$-$088, WD\,1132$-$325 and WD\,1917$+$386. In total, we identify five stars in the sample for which metals are only detected in the ultraviolet. Overall, we find that 30\,per\,cent of the 13\,pc white dwarfs show spectroscopic evidence of evolved planetary systems. Our analysis reveals no measurable difference between the hydrogen content of DQ and DC white dwarfs, although the upper limits of carbon in DCs are significantly below that of the DQ population. We find a multiplicity fraction of 33\,per\,cent for the 13\,pc white dwarfs.
\end{abstract}



\section{Introduction}

White dwarfs are the evolutionary endpoint of 95\,per\,cent of all stars. Despite the recent discoveries of hundreds of thousands of white dwarf candidates with the \textit{Gaia} mission \citep{Gaia2021,Gentile2021}, even those in our immediate Solar neighbourhood remain incompletely characterised across all wavelengths. The white dwarfs in the local volume exhibit a wide diversity of masses, ages, atmospheric compositions (including evidence of evolved planetary systems), multiplicity, and magnetic field strengths. Volume-limited studies provide a critical insight into white dwarf cooling rates, evolution of planetary systems, and the roles of binarity and magnetic fields in stellar evolution \citep{Holberg2002,Holberg2008,Giammichele2012,Toonen2017,Hollands2018_Gaia,Bagnulo2022,Caron2023,OBrien2024,Kilic2025_100pc,OuldRouis2026}. 

Optical spectroscopy of white dwarfs is commonplace, and enables an initial characterisation of atmospheric properties including chemical composition. Ultraviolet (UV) observations provide new insights into a star. Some photospheric metal absorption lines that are not visible in the optical appear in the UV \citep{Weidemann1980,Koester2000,Gaensicke2012,Jura2012,Sahu2024_dust}, enabling the characterisation of the geochemistry of accreted planetesimals \citep{Koester2014} or constraining the convective dredge-up of carbon from the interior \citep{Bedard2024,Sahu2025_NatAs}. In addition to chemical composition, the UV continuum flux enables models to be fitted over a larger wavelength range to better constrain the stellar parameters such as mass, surface gravity ($\log [g$(cm s$^{-2})]$), effective temperature (\Teff) and age \citep{Lajoie2007,Sahu2023,Elms2024}. 

Most previous spectroscopic UV white dwarf surveys have focused on warmer ($T_{\rm eff} \gtrsim 10\,000$\,K) objects that are bluer, brighter and have more lines in this wavelength range. However, white dwarfs warmer than 10\,000\,K represent just 15\,per\,cent of the local volume \citep{OBrien2024}. The white dwarf temperature distribution in the local volume is primarily a function of white dwarf cooling rates, which map the population back to the underlying convolution of the initial mass function and the star formation history of the Milky Way. Cooler and older white dwarfs emit less flux in the near-UV, and as such have not been studied extensively at UV wavelengths. Spectroscopic UV observations of cool white dwarfs are key to test various aspects of white dwarf physics and metal accretion \citep{Wolff2002,Saumon2014,Hollands2022}

The high-density atmospheres of the coolest white dwarfs are notoriously challenging to model \citep{Saumon1999,Bergeron2001}, particularly regarding collision induced absorption (CIA) in the infrared \citep{Blouin2017, Elms2022, Bergeron2022,Saumon2022,Blouin2024}. The accurate modelling of cool white dwarf atmospheres is a significant issue that affects most of the local white dwarf population, as over half of local white dwarfs are cooler than 6000\,K \citep{OBrien2024}. The Ly\,$\alpha$ line is collision-broadened into the blue part of the optical, as demonstrated by improved Ly\,$\alpha$ broadening profiles \citep{Kowalski2006,Allard2009,Saumon2014,Sahu2025}. Line profiles used in the modelling perform well \citep{Saumon2014}, and this broadening is caused mostly by H--H$_{2}$ and H--He collisions. However, even with the well tested and validated Ly\,$\alpha$ opacities for
the UV and optical, and CIA opacities in the infrared, there are still subtle discrepancies between the predicted and observed photometry and spectral energy distributions, visible in analysis of large data sets \citep{OBrien2024}.

In this work, we present a detailed analysis of UV spectroscopy from the \textit{Hubble Space Telescope} (\textit{HST}) of white dwarfs within 13\,pc of the Sun, some of our nearest stellar neighbours. Given their proximity, white dwarfs within 13\,pc are among the brightest members of their spectral types, providing uniquely high-quality data for atmospheric modelling. The 13\,pc UV observations are supplemented with medium-resolution optical spectroscopy. We combine the STIS data, which is flux-calibrated up to the 1\,per\,cent level \citep{Bohlin2019}, with optical and IR photometry from \textit{Gaia} Data Release 3 (DR3) \citep{Gaia2021}, the Two Micron All-Sky Survey (2MASS) \citep{2MASS2006} and the Wide-field Infrared Survey Explorer (WISE) \citep{WISE2010}. We then perform spectrophotometric fits from the UV to the IR incorporating the STIS spectra plus the photometry, as well as fits to spectral features to determine atmospheric compositions. 

By fitting multi-wavelength flux-calibrated data across the UV-to-IR range, there is an immediate visual insight into discrepancies and offsets between the fits to different regions of the white dwarf continuum. In this work, we provide best-fitting parameters for every 13\,pc white dwarf, and discuss the differences between these fits in different wavelength regimes. The UV spectra enable us to determine precise limits on carbon and hydrogen in DC white dwarfs and hydrogen in DQ white dwarfs. We also test the impact of fitting magnetic DA white dwarfs with radiative models instead of convective models. We discuss the binary population and photometric variability within the 13\,pc sample, and additionally provide insight into the geology of the material accreting onto 13\,pc white dwarfs.

\section{Observations}

\begin{figure}
    \centering
	\includegraphics[width=\columnwidth]{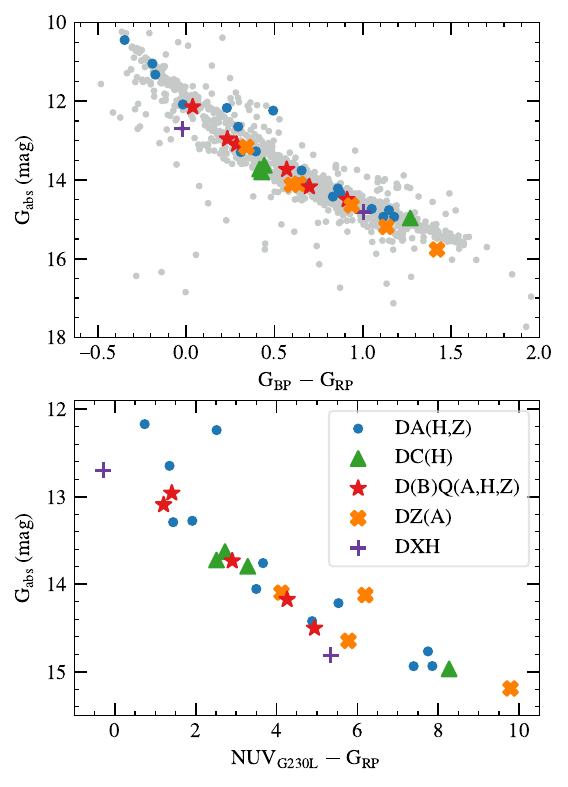}
	\caption{HR diagrams for the 13\,pc white dwarf sample, separated by spectral type. \textit{Top}: A \gaia\ HR diagram using the G, G\textsubscript{BP} and G\textsubscript{RP} bands. The grey points are white dwarfs within 40\,pc \citep{OBrien2024}. \textit{Bottom}: A \gaia\ HR diagram with newly created NUV\textsubscript{G230L} photometric points. The NUV\textsubscript{G230L} colour was generated by convolving the STIS G230L spectra with the GALEX NUV filter. Only white dwarfs with G230L observations are shown in this lower plot.}
    \label{fig:plot_HR}
\end{figure}

There are 44 spectroscopically confirmed white dwarfs known to be within 13\,pc of the Sun \citep{OBrien2024}. These reside in 42 stellar systems, comprising 28 isolated white dwarfs and 14 multiple-star systems containing at least one white dwarf (of which two are WD\,$+$\,WD binaries). The details of the observations of the white dwarfs that were analysed in this work are listed in Table~\ref{tab:13pc_observations}. \textit{HST} observations of 36 targets within 13\,pc were conducted during Cycle~23 (program 14076; \citealt{Gaensicke2015}), with the Space Telescope Imaging Spectrograph (STIS) and the Cosmic Origins Spectrograph (COS). All of the Cycle~23 observations were completed between November 2015 and January 2017.

WD\,0727$+$482\,A$+$B is a white dwarf $+$ white dwarf binary with a separation of 0.66\,arcsec. For the Cycle~23 STIS G230L observation, both stars were positioned along the 52"$\times$2" long-slit, and were later extracted individually. Shorter slits were used for three of the Cycle~23 STIS G230L observations to limit contamination from a nearby companion star. WD\,0208$-$510 is 2.1\,arcsec away from a K dwarf (GJ\,86\,A), and was observed with the 52"$\times$0.2" aperture. WD\,0426$+$588 is 9.0\,arcsec away from an M dwarf (Stein\,2051\,A), and was observed with the 52"$\times$0.5" aperture. WD\,1132$-$325 is 15\,arcsec  away from a K dwarf (GJ\,432\,A), and was also observed with the 52"$\times$0.5" aperture. Additional archival \textit{HST} observations were compiled in this work to improve the completeness of the 13\,pc sample. Details of program IDs are in Table~\ref{tab:13pc_observations}.

Eighteen white dwarfs within 13\,pc were observed with the X-shooter spectrograph on the VLT \citep{Vernet2011}. The data were reduced using the \textsc{reflex} pipeline \citep{Freudling2013}, and we performed telluric corrections using \textsc{molecfit} \citep{Kausch2015,Smette2015}. Four white dwarfs have archival UVES data, and reduction was carried out using the same procedure. Six white dwarfs were observed with the ISIS instrument on the WHT. All of the WHT ISIS observations were taken on 2016 September 04. Some additional optical spectra were taken from the Montr\'eal White Dwarf Database \citep{Dufour2017}, and the references to the original publications of these spectra are found in Table~\ref{tab:13pc_observations}.

\begin{table*}
    \centering
        \caption{Details of UV and optical spectroscopy available for 13\,pc white dwarfs.}
    \begin{tabular}{lllll}
         \hline
         WD & \textit{HST} UV&\textit{HST} UV& Optical observation &Optical\\
         name &  observations&PID& or citation &PID\\
        \hline
         0009$+$501 & STIS G230L 52"$\times$2"&14076&WHT/ISIS &--\\
         0038$-$226 &STIS G230L 52"$\times$2"&14076&VLT/X-shooter & ESO 097.D$-$0064(A)\\
         0046$+$051 &STIS G230L 52"$\times$2"&14076&VLT/X-shooter &ESO 098.D$-$0392(A)\\
         0135$-$052 &STIS G230L 52"$\times$2"&14076&VLT/UVES &ESO 165.H$-$0588(A)\\
         0141$-$675 &STIS G230L 52"$\times$2"&14076&VLT/X-shooter & ESO 099.D$-$0661(A)\\
         0208$-$510& STIS G230L 52"$\times$0.2"&14076&\textit{HST}/STIS G430L 52"$\times$0.2"&\textit{HST} 12548\\
         0245$+$541 &STIS G230L 52"$\times$2"&14076& \citet{Zuckerman2003}&--\\
         0310$-$688 &STIS E230M/E140M 0.2"$\times$0.2"&14076&VLT/X-shooter &ESO 1103.D$-$0763(C)\\
         0413$-$077 &STIS E230M/E140M 0.2"$\times$0.2"&14076&VLT/X-shooter &ESO 098.D$-$0392(A)\\
         0426$+$588&STIS G230L 52"$\times$0.5"; COS G140L&14076&\citet{Bergeron1997} &--\\
         0435$-$088 &STIS G230L 52"$\times$2"&14076&VLT/X-shooter &ESO 098.D$-$0392(A)\\
         0548$-$001 &STIS G230L 52"$\times$2"&14076&VLT/X-shooter &ESO 099.D$-$0661(A)\\
         0552$-$041 &STIS G230L 52"$\times$2"&14076&VLT/X-shooter &ESO 098.D$-$0392(A)\\
         0553$+$053 &STIS G230L 52"$\times$2"&14076&VLT/X-shooter &ESO 098.D$-$0392(A)\\
         0642$-$166& STIS E230H/E140H 0.2"$\times$0.09"&12981& \textit{HST}/STIS G430L/G750M 52"$\times$2"&\textit{HST} 12606\\
         0727$+$482\,A& STIS G230L 52"$\times$2"&14076&\citet{Bergeron2001} &--\\
         0727$+$482\,B& STIS G230L 52"$\times$2"&14076&\citet{Bergeron2001} &--\\
         0736$+$053& STIS G230LB  52"$\times$0.2"&7398& \textit{HST}/STIS G430L 52"$\times$2"&\textit{HST} 7398\\
         0738$-$172& FOS G270H &2593&\citet{Giammichele2012} &--\\
         0752$-$676 &STIS G230L 52"$\times$2"&14076&VLT/X-shooter &ESO 098.D$-$0392(A)\\
         0810$-$353& -- &--&\citet{Bagnulo2020} &--\\
         0821$-$669 &STIS G230L 52"$\times$2"&14076&\citet{Subasavage2007} &--\\
         0839$-$327 &STIS G230L 52"$\times$2"; COS G140L&14076&VLT/X-shooter &ESO 098.D$-$0392(A)\\
         0912$+$536 &STIS G230L 52"$\times$2"; COS G140L&14076& \citet{Bergeron1997} &--\\
         1055$-$072 &STIS G230L 52"$\times$2"&14076&VLT/X-shooter &ESO 1103.D$-$0763(C)\\
         1132$-$325 &STIS G230L 52"$\times$0.5" &14076&\citet{Giammichele2012} &--\\
         1142$-$645 &STIS G230L 52"$\times$2"; COS G140L&14076&VLT/X-shooter & ESO 099.D$-$0661(A)\\
         1202$-$232 &STIS G230L 52"$\times$2"; COS G140L&14076&VLT/X-shooter &ESO 098.D$-$0392(A)\\
         1334$+$039 &STIS G230L 52"$\times$2"&14076&VLT/X-shooter &ESO 099.D$-$0661(A)\\
         1345$+$238 &STIS G230L 52"$\times$2"&14076&\citet{Koester2001} &--\\
         1620$-$391 &STIS E230M/E140M 0.2"$\times$0.2"&14076&VLT/UVES & ESO 167.D$-$0407(A)\\
         1630$+$089&--  &--&\citet{Limoges2013} &--\\
         1647$+$591 & COS G130M/G160M &14076&WHT/ISIS &--\\
         1748$+$708 &STIS G230L 52"$\times$2"&14076&WHT/ISIS &--\\
         1900$+$705 &STIS G230L 52"$\times$2"; COS G140L&14076&WHT/ISIS &--\\
         1917$-$077 &STIS E230M 0.2"$\times$0.2"; COS G130M/G160M&14076&VLT/X-shooter & ESO 4109.A$-$0015(A)\\
         1917$+$386 &STIS G230L 52"$\times$2"&14076&WHT/ISIS &--\\
         1953$-$011 &STIS G230L 52"$\times$2"&14076&VLT/UVES & ESO 165.H$-$0588(A)\\
         2140$+$207 &STIS G230L 52"$\times$2"; COS G140L&14076& VLT/UVES &ESO 097.D$-$0063(A)\\
         2150$+$591 & -- &--& \citet{Tremblay2020} &--\\
         2251$-$070 &-- &--&VLT/X-shooter &ESO 098.D$-$0392(A)\\
         2359$-$434 &STIS G230L 52"$\times$2"&14076&VLT/X-shooter &ESO 099.D$-$0661(A)\\
         G\,203-47\,B& STIS G230L 52"$\times$2"&17778& --& --\\
         \hline
    \end{tabular}
    \label{tab:13pc_observations}
\end{table*}

\begin{table}
    \centering
        \caption{All spectroscopically confirmed white dwarfs within 13\,pc of the Sun, ordered by distance. Distances were determined using \gaia\ DR3 parallaxes of the white dwarfs or their companion stars in all cases, except for Procyon\,B \citep{vanLeeuwen2007}. SpT refers to the spectral type. Comp. refers to the dominant atmosphere species.}
    \begin{tabular}{lllll}
         \hline
         WD name& Alternative name&SpT  &Comp.&Distance  \\
          &  &  &&(pc)  \\
        \hline
         0642$-$166& Sirius\,B&DA  &H&2.67 \\
         0736$+$053& Procyon\,B&DQZ  &He&3.51 \\
         0046$+$051 &van\,Maanen\,2&DZA  &He&4.31  \\
         1142$-$645 &LAWD 37&DQ  &He&4.64  \\
         0413$-$077 &40 Eri B&DA  &H&5.01  \\
         0426$+$588&Stein 2051 B&DC  &He&5.52  \\
         1748$+$708 &G240-72&DXH  &--&6.21  \\
         0552$-$041 &LP 658-2&DZ  &He&6.44  \\
         --& G203-47\,B& --&--&7.60\\
         0553$+$053 &G99-47&DAH  &H&8.12  \\
         0752$-$676 &L97-12&DA  &H&8.17  \\
         2359$-$434 &GJ 915&DAH  &H&8.33  \\
         1334$+$039 &Wolf 489&DA  &H&8.35  \\
         2150$+$591 & --&DAH  &H&8.46  \\
         0839$-$327 &GJ 318&DA  &H&8.52  \\
         2251$-$070 &LP 701-29&DZ  &He&8.54  \\
         0038$-$226 &LHS 1126&DQpec  &He&9.10  \\
         0738$-$172& LAWD 25&DZA  &He&9.15 \\
         0435$-$088 &LHS 194&DQZ  &He&9.40  \\
         1132$-$325 &GJ 432 B&DZ  &He&9.56  \\
         0141$-$675 &L88-59&DAZ  &H&9.72  \\
         0912$+$536 &G195-19&DCH  &He&10.27  \\
         0310$-$688 &LB 3303&DAZ  &H&10.40  \\
         1202$-$232 &LP 852-7&DAZ  &H&10.43  \\
         1917$-$077 &LAWD 74&DBQA  &He&10.51  \\
         0821$-$669 &SCR 0821$-$6703&DA  &H&10.68  \\
         0208$-$510& GJ 86 B&DQ  &He&10.80  \\
         0245$+$541 &LHS 1446&DAZ  &H&10.87  \\
         0009$+$501 &LHS 1038&DAH  &H&10.87  \\
         1647$+$591 &G226-29&DAV  &H&10.95  \\
         2140$+$207 &LHS 3703&DQ  &He&11.04  \\
         0810$-$353& -- &DAH &H& 11.17 \\
         0548$-$001 &G99-37&DQH  &He&11.22  \\
         0727$+$482\,A& G107-70A& DA  &H&11.27\\
         0727$+$482\,B& G107-70B& DA  &H&11.27\\
         1953$-$011 &LHS 3501&DAH  &H&11.57  \\
         1345$+$238 &LP 380-5&DC  &H&11.86  \\
         1917$+$386 &GJ 1234&DZ  &He&11.87  \\
         1055$-$072 &LHS 2333&DC  &He&12.27  \\
         0135$-$052 &L870-2&DA$+$DA   &H&12.62  \\
         1900$+$705 &Grw$+$70$^{\circ}$8247&DXH  &--&12.88  \\
         1620$-$391 &CD-38 10980&DAZ  &H&12.91  \\
         1630$+$089& G138-38&DA  &H&12.93 \\
         \hline
    \end{tabular}
    \label{tab:13pc_info}
\end{table}

Three confirmed white dwarfs within the 13\,pc volume do not have \textit{HST} UV spectra available. WD\,0810$-$353, WD\,1630$+$089 and WD\,2150$+$591 were discovered to be within 13\,pc by \gaia, years after the Cycle~23 program, and have therefore not yet been observed in the UV with \textit{HST}. WD\,2251$-$070 was observed with STIS G230L over three orbits as part of the Cycle~23 program but it was so faint in the UV that there was no detection. The white dwarf in the post-common envelope binary system G\,203-47 was directly detected using \textit{HST} STIS G230L and was found to be $\approx$\,5300\,K \citep{OBrien2026}. The UV spectrum of G\,203-47 is not analysed further in this work.

All 13\,pc white dwarfs analysed in this work were checked for magnetism using sensitive spectropolarimetric measurements, as part of the larger 20\,pc volume sample \citep{Bagnulo2021}. All spectral types in Table~\ref{tab:13pc_info} account for this magnetic field search, with an ‘H' at the end of the spectral type indicating a magnetic field detection of any kind. Additionally, eight members of the 13\,pc sample were observed at a resolution of 34\,000 around the Ca\,\textsc{ii}~H and K doublet, providing calcium abundance detections and upper limits for H-atmosphere white dwarfs \citep{Zuckerman2003}. The remaining members of the sample could have weak metal features and should be followed up with a high-resolution optical spectrograph. The spectral types in Table~\ref{tab:13pc_info} are based on the \citet{Sion1983} naming system, and are up-to-date following our analysis. For simplicity, due to our multi-wavelength dataset, we opted to combine UV and optical features for an overall spectral type classification. 

\section{Model atmospheres and fitting}
\label{sec:models}

The coolest white dwarfs within 13\,pc have almost no flux in the STIS G230L band, except for at the reddest end. When fitted in isolation with atmosphere models, these cool STIS spectra did not provide meaningful white dwarf parameters, and as such, additional optical and IR photometry was required to constrain their parameters. For this reason, we employed a universal `combined' fitting method, where STIS spectra, together with \textit{Gaia} DR3 $G$, $G_\mathrm{BP}$, $G_\mathrm{RP}$, 2MASS $J, H, K_\mathrm{s}$ and WISE $W1$, $W2$ photometry were combined into an overall spectrophotometric fit. This fit enabled us to obtain atmospheric parameters, including \Teff\ and \logg, as well as mass and radius. This method constrained parameters of the coolest members of the sample well, and the fit was not weighted in favour of any particular dataset. 

The STIS spectra are flux-calibrated at the 3\,per\,cent level \citep{Elms2024}, and \textit{Gaia} DR3 $G$, $G_\mathrm{BP}$, $G_\mathrm{RP}$, and 2MASS $J, H, K_\mathrm{s}$ and WISE $W1$, $W2$ photometry are also well flux-calibrated. This flux calibration enables a precise full continuum fit. We elected not to incorporate Pan-STARRS $grizy$ and SDSS $ugriz$ optical photometry in the fits, because many of the 13\,pc white dwarfs were so bright that they saturated in multiple Pan-STARRS and SDSS filters, and we wanted a uniform sample of uncontaminated photometry. Additionally, \gaia\ has proven to be as precise and accurate as Pan-STARRS and SDSS for local white dwarfs \citep{OBrien2024}. The parallax of each star was fixed to the \textit{Gaia} DR3 value during the fit. We assumed the distance to be the inverse of the parallax, which is appropriate for these sources as their \gaia\, \texttt{parallax\_over\_error}\,$>$\,1800, and therefore they have a negligible measurement error \citep{BailerJones2015}. 

The STIS E140M and E230M observations initially suffered from echelle order ripples, which were corrected using the \textsc{stisblazefix} package before fitting \citep{Baer2018}. The atomic line data used to identify spectral features in this work were obtained from NIST \citep{Kramida2018}. Photospheric lines (from accreted metals) and interstellar lines (from gas in the line of sight) are usually offset due to the white dwarf gravitational redshift, and the STIS E140M/H and E230M/H gratings have a high enough resolution that the narrow photospheric lines and interstellar lines are resolved. The line components of Si\,\textsc{ii} 1265\,\AA\ have a lower energy level of 0.035\,eV, corresponding to $\approx$\,400\,K. These levels are not populated in the cool interstellar medium, and so all lines with the same radial velocity as the Si\,\textsc{ii} 1265\,\AA\ line must be photospheric. Interstellar lines were identified and then fitted with a Gaussian using a least squares fitting method, or in the case of the two components in WD\,0642$-$166 (Sirius\,B), two Gaussians. The regions of the interstellar lines were then masked from the spectra prior to fitting. 

\subsection{H-atmosphere fitting}
\label{sec:Hatmosphere}

We fitted H-atmosphere white dwarf data with updated 3D local thermodynamic equilibrium (LTE) model spectra\footnote{\url{https://warwick.ac.uk/fac/sci/physics/research/astro/people/tremblay/modelgrids/}} \citep{Tremblay2013, Tremblay2015}. These models include Stark broadening profiles \citep{Tremblay2009}, neutral line broadening, charged- and neutral-particle non-ideal effects \citep{Hummer1988}, and H\textsubscript{2} molecular lines. The models also incorporate the Ly\,$\alpha$ quasistatic single-perturber H–H\textsubscript{2} red-wing opacity for $T_{\rm eff} \leq 9000$\,K \citep{Kowalski2006}, and the multi-perturber model Ly\,$\alpha$ H--H and H--H$^{+}$ quasi-satellites applied at $T_{\rm eff} > 9000$\,K \citep[][via an updated version\footnote{\url{https://www.iap.fr/useriap/allard/lymantables.html}}]{Allard2009}. The model spectra rely on the collision-induced absorption H$_2$-H$_2$ opacity \citep[][and from the Hitran database\footnote{\url{http://hitran.org/cia/}}]{Borysow2001}, including high-density corrections \citep{Hare1958}. These ingredients allowed us to fit Ly\,$\alpha$ and the continuum from UV-to-IR of pure-H atmosphere white dwarfs. The models with H\textsubscript{2} molecular lines switched on were only used when fitting the high-resolution COS spectra. Prior to fitting, the models were convolved with the appropriate line spread function for the required STIS or COS grating\footnote{\url{https://www.stsci.edu/hst/instrumentation/stis/performance/spectral-resolution}}, using the \textsc{linetools Python} package\footnote{\url{https://github.com/linetools/linetools}}.

The overabundance of the $\rm H_3^+$ species in cool H-atmosphere models has recently been identified as a likely reason for the disagreement between models and \gaia\ data for white dwarfs cooler than 6000\,K \citep{Kowalski2026}. Following the work of \citet{Kowalski2026}, it was realised that the H-atmosphere models had to be updated to remove the contribution of nuclear spin degeneracy to the \citet{Neale1995} $\rm H_3^+$ partition function (S. Blouin private communication; see also \citealt{Kilic2026})\footnote{The nuclear spin degeneracy is not included when considering partition functions of other hydrogen species, resulting in $\rm H_3^+$ being treated inconsistently in previous models.}, which had previously led to the overestimation of the partition function and abundance of $\rm H_3^+$ in the models. We applied this correction to the models below 6000\,K, resulting in a reduction of the H$^-$ opacity.

The grids of pure-H atmosphere model spectra that we used for fitting spanned \Teff\,$=$\,1500\,$-$\,40\,000\,K, and \logg\,$=$\,7.0\,$-$\,9.0 in steps of 0.5 dex. LTE models are appropriate for fitting the white dwarfs in this sample because they are well below 40\,000\,K, and non-LTE effects only become significant above that temperature, or in line cores at high-resolution. For the mass-radius relation, we adopted a thick H-layer, $q_\mathrm{H} = M_\mathrm{H}/M_\mathrm{WD} = 10^{-4}$ \citep{Bedard2020}. 

For the combined (STIS plus photometry) fitting of DA white dwarfs, we expanded on the fitting routine presented in \citet{Sahu2023}. Briefly, the STIS spectra and photometry were combined, then compared to pure-H model spectra using the reduced $\chi^2$ metric, and then the data were fitted using the least-squares method. Filter response functions and zeropoint fluxes for fitting photometry were taken from the Spanish Virtual Observatory (SVO) Filter Profile Service \citep{Rodrigo2012, Rodrigo2020}. The SVO zeropoints were derived using \texttt{alpha\_lyr\_stis\_010.fits}\footnote{\url{https://archive.stsci.edu/hlsps/reference-atlases/cdbs/calspec/alpha_lyr_stis_010.fits}} as the Vega reference spectrum for $\lambda < 1 \mu$m, and \texttt{alpha\_lyr\_stis\_011.fits}\footnote{\url{https://archive.stsci.edu/hlsps/reference-atlases/cdbs/calspec/alpha_lyr_stis_011.fits}} for $\lambda > 1 \mu $m.

Where multiple Balmer lines were visible in an optical spectrum, they were independently fitted using a \textsc{python} implementation of a fitting code that was created as part of the 4MOST white dwarf fitting pipeline \textsc{4MOST\_WDpipe}\footnote{\url{https://github.com/NPGFusillo/4MOST_WDpipe}}. For these fits, we used pure-H 1D LTE models \citep{Tremblay2011} (mixing length calibration of ML2/$\alpha$ = 0.8) with 3D corrections \citep{Tremblay2013}. In cool DA white dwarfs, hydrogen line broadening is increasingly dominated by neutral interactions, with resonance broadening as the main contributor, traditionally treated using the parameters of \citet{Ali1965}. Modelling dense, non-ideal plasmas is challenging, as volume effects from finite particle sizes can induce pressure ionisation. Most atmosphere codes adopt the Hummer–Mihalas occupation probability formalism, which treats atoms as hard spheres and removes states when interparticle distances fall below their radii \citep{Hummer1988}. However, a straightforward application of this model yields systematically low $\log g$ values below $T_{\mathrm{eff}} \lesssim 8000$ K, where neutral interactions dominate. To mitigate this, the hydrogen radius is reduced by a factor of two for the higher Balmer lines \citep{Tremblay2010}. We did not fit the Balmer lines and photometry simultaneously, as there is a well-known offset of $\approx$2\,per\,cent between photometric and spectroscopic \Teff\ solutions for DA white dwarfs that is present when fitting different homogeneous spectroscopic and photometric data sets \citep{Tremblay2019,Tremblay2020,Genest-Beaulieu2019,Cukanovaite2021,Sahu2023,OBrien2023}. 

For the confirmed unresolved double white dwarf binary system in the 13\,pc sample, WD\,0135$-$052, we fitted the Balmer lines from UVES spectra \citep{Napiwotzki2020} to determine the parameters of both components of the system. We used the code \textsc{wd$-$bass}\footnote{\url{https://github.com/JamesMunday98/WD-BASS}} to fit the Balmer lines and photometry \citep{Munday2024}, where the same updated models as discussed above were used for fitting \citep{Tremblay2013, Tremblay2015}. \textsc{wd$-$bass} is a specialist code for fitting unresolved double white dwarf systems. It is standard to incorporate photometry from SDSS in fits with \textsc{wd$-$bass} as well as \gaia\ photometry, and we removed the saturated SDSS $u$ and $z$ points prior to fitting. There were degeneracies in the fit due to the two white dwarfs having similar \Teff, so we restricted the \Teff\ bounds to remove the issue of stuck walkers in the MCMC.

We obtained the \Teff\ and \logg\ of the metal-enriched H-rich atmosphere white dwarfs (DAZ) using the combined fitting method described above, masking the metal lines during the fit as they were sufficiently narrow to not significantly impact the overall shape of the SED. Furthermore, the metal abundances were low enough to have negligible influence on the equation-of-state and any hydrogen opacity including line broadening. Metal abundances were then derived using updated hydrogen-rich white dwarf atmosphere models based on the \citet{Koester2010} code. Fixing the atmospheric parameters (\Teff, \logg) to the values obtained from the combined fit, the model grids were computed for each star over the abundance range $-13 < \log \mathrm{(Z/H)} < -6$ in steps of 0.2\,dex. Calcium abundances were derived from the optical spectra by normalising the continuum in the region surrounding the Ca\,\textsc{ii}\,K line and comparing the observed profiles with the model grid using a $\chi^2$ minimisation technique. For carbon, oxygen, magnesium, and silicon, we used the flux-calibrated STIS spectra and performed fits in the wavelength regions encompassing the absorption features of each element. As with the combined fits to the DA white dwarfs, the synthetic spectra were convolved with the appropriate STIS line spread functions corresponding to the instrumental configuration of each observation, using the \textsc{linetools} python package. We then followed the methodology described in \citet{Williams2025} to model interstellar absorption simultaneously with the photospheric lines using the convolved spectra, thereby deriving the best-fitting elemental abundances and their corresponding uncertainties.


\subsection{He-atmosphere fitting}
\label{sec:Heatmosphere}

Our analysis of helium-rich white dwarf spectra used models calculated with the code of D.~Koester. The basic algorithms and atomic data are described in \citet{Koester2010}. The code and data have been updated continuously since then, with details occasionally described in the literature \citep{Koester2014,Hollands2017,Gaensicke2018,Hollands2022,Elms2022}. Here we summarise the most important changes to the code that are relevant for the cool, helium-rich white dwarfs in the 13\,pc sample.

\noindent {\bf Equation of state:} Deviations from the ideal gas law become noticeable at densities of 0.01\,g cm$^{-3}$, and significant for values larger than 0.1\,g cm$^{-3}$. The latter value is reached in the photosphere of pure-He white dwarfs with \Teff\ $\approx $ 7500\,K, and for somewhat lower \Teff\ values if traces of elements with lower ionisation potentials (hydrogen or metals) are present. For these objects, non-ideal effects in the equation of state and absorption coefficients must be taken into account. Since a fully consistent treatment that can be applied to stellar atmosphere calculations is not available, simplifying approximations are used.

The most important quantity for the equation of state is the ionisation balance of helium, since it determines the He$^-$ absorption coefficient. \citet{Kowalski2007} calculated this degree of ionisation as a function of density for two temperatures: 11\,605\,K (1\,eV) and 5803\,K (0.5\,eV). Their calculations of the electron interaction energy, which is an important factor for determining the lowering of the ionisation potential, show that it is roughly linear with density (their Fig. 3) and exhibits only a weak temperature dependence at these specific temperatures (their Fig. 4). However, the overall ionisation balance and band gap become highly sensitive to temperature as it increases, particularly above 1.5\,eV (their Fig. 7). At these higher temperatures, the atmospheric density is much lower. A correction term for the free energy can be constructed, which scales as $\approx N^2/V$, where $N$ is the number of helium atoms and $V$ is the volume. The change of the ionisation potential derived in the usual way from that correction term is thus linear in density. The constant in the relation was determined from a comparison with the result from \citet{Kowalski2007}. With this procedure the ionisation balance from the code closely follows the result of the much more complicated calculation of \citet{Kowalski2007}. Applying a correction term to the free energy enables the derivation of a consistent correction to pressure and the secondary quantities of specific heat and adiabatic gradient. The pressure correction becomes large when the density reaches above 1\,g cm$^{-3}$, and is very similar to the calculations of \citet{Becker2014} (their Fig. 5). It should be noted that this treatment of the helium ionisation is similar to the so-called ``excluded volume'' model of the free energy. We note that the recent analysis of multi-wavelength spectra of cool, carbon atmosphere white dwarfs in \citet{Farihi2026} allowed better constraints of the highly uncertain ionisation equilibrium in dense helium. The authors applied a scaling factor to the ionisation equilibrium of \citet{Kowalski2007}, which indicated that the results of advanced quantum-mechanical calculations given in Fig. 10 in that paper are closer to reality. As the problem of ionisation equilibrium in condensed, helium atmospheres requires further investigation, we have not updated our models in view of the recent findings, but comment on differences between the models when discussing DQ stars.

\citet{Blouin2018b} calculated the lowering of the ionisation potentials for five important metals: Na, Mg, Fe, Ca, and C. We used the equations from \citet{Blouin2018b}, which depend on density and weakly on temperature, for our calculations. For the calculation of line spectra the non-ideal changes to the occupation numbers of all states are important, which is treated with the occupation probability formalism of \citet{Hummer1988}. However, it is known that this formalism does not lead to pressure ionisation of hydrogen and helium \citep{Mihalas1990}, which therefore has to be treated separately as described above.

\noindent {\bf Continuous absorption coefficients:} The most important continuous absorption coefficients in cool He-rich atmosphere white dwarfs are the free-free absorption of He$^-$ and He Rayleigh scattering.  High-density corrections for He$^-$ have been determined by \citet{Iglesias2002}, and were recalculated by S. Blouin (priv. comm.). The high-density Rayleigh scattering coefficients are from \citet{Rohrmann2018}. At low temperatures in He-rich models, CIA becomes important. He$-$H$_2$ \citep{Abel2012} and He$-$He$-$He are incorporated in the models \citep{Kowalski2014}. If the abundance of metals is large (e.g. in DQs), the photoabsorption cross sections from \textsc{TOPbase} are included \citep{Seaton1995}.

\noindent {\bf Spectral line absorption:} At high densities, many spectral lines show broad profiles with asymmetries, and sometimes ``satellites'' on their wings that are not consistent with the classical Voigt profiles of the absorption coefficients. These lines are analysed with ``unified theories'' that aim to describe the core as well as the far wings of the lines, including possible satellites \citep{Allard1982,Allard1999}. We use an independent method with improved numerical algorithms. Such line profiles are now available for Ly\,$\alpha$ (broadened by H, H$^+$, H$_2$, He), Ly~$\beta$ (H, H$^+$), Ly~$\gamma$ (H, H$^+$), C\,\textsc{i}\,1657, 1931, 2479, Na\,\textsc{i}\,5892, 5898, Mg\,\textsc{i}\,2853, 5169, 5174, 5185, Mg\,\textsc{ii}\,2796, 2804, Ca\,\textsc{i}\,4228, and Ca\,\textsc{ii}\,3935, 3870\,\AA.

The line lists of the C$_2$ and CH bands from the \textsc{ExoMol} database have been implemented in the code for DQ white dwarfs \citep{Tennyson2016,Tennyson2024}. Using the ``Just Overlapping Line'' formulation, as developed in \citet{Zeidler1982} gives almost identical results with much less computational effort, so that formulation was used for C$_2$ in the code. The shift of the electronic energy levels, that causes the shift of the Swan bands in the DQpec stars, was found by quantum mechanical simulations to be $d$\,(eV)\,$=a\cdot\rho$\,(g cm$^{-3}$) \citep{Kowalski2010}. However, \citet{Kowalski2010} and \citet{Blouin2019b} changed the constant to 0.2, for better fits to the observations, and in some cases a value of 0.1 gave the best fits. As noted by \citet{Blouin2019b}, no single value of this constant fits all objects. We tested the impact of this parameter on our objects and other DQ white dwarfs, and settled on 0.1 for the model grids. These problems are also related to uncertainties in helium ionisation equilibrium and hydrogen abundance \citep{Kowalski2010}. Recently, with better constraints on the ionisation equilibrium, \citet{Farihi2026} obtained consistent fits to a series of cool DQ stars. Determining a precise constraint is not a topic of this work, and distortion of C$_2$ bands does not affect the band strength and derived carbon abundances substantially. 

\subsubsection{DC fitting}
\label{sec:DC_fitting}

We fitted three DC white dwarfs with a similar combined fitting method as described in Section~\ref{sec:Hatmosphere}. He-atmosphere models with trace hydrogen were generated using the \citep{Koester2010} code. The He $+$ H grids of model spectra spanned \Teff\,$=3000-20\,000$\,K in steps of 250\,K, \logg\,$=7.0-9.5$ in steps of 0.25\,dex, and $\log$(H/He) $=-6.0$ to 0.0 (plus pure-He) in steps of 0.25 dex. We chose to apply He $+$ H models to fit the combined STIS spectra and photometry of the featureless DC white dwarfs, WD\,0426$+$588, WD\,1055$-$072, and WD\,0912$+$536, as their \Teff\ values are high enough that Balmer lines would be detectable in the spectra if they had pure-H atmospheres. We applied a reduced $\chi^2$ least-squares fit to the STIS spectra and photometry, to determine \Teff\ and \logg. For the mass-radius relation, we used a thin H-layer option, $q_\mathrm{H} = 10^{-10}$ \citep{Bedard2020}. Since hydrogen is an additional free parameter in the model spectra, we interpolated over the grids to find the best solution for $\log$(H/He) as well as \Teff\ and \logg. Once the initial fit was complete, we generated grids of models incorporating varying levels of trace carbon, with all other parameters fixed, to determine upper limits on the carbon abundances. 

\subsubsection{DQ fitting}
\label{sec:DQ_fitting}

To fit the DQ white dwarfs, He-atmosphere models with carbon were generated using the \citep{Koester2010} code. The He $+$ C grids of model spectra spanned \Teff\,$=3000-12\,000$\,K in steps of 250\,K, \logg\,$=7.25-9.75$ in steps of 0.25 dex, and $\log$(C/He)$=-3.0$ to $-$9.5 in steps of 0.5 dex. Some specific objects required tailored grids with hydrogen or metals included, which are discussed further in Section~\ref{sec:dq_fits}.

In the STIS spectra, the strong neutral atomic carbon lines, C\,\textsc{i} 1657\,\AA, 1930\,\AA\ and 2478\,\AA, were substantially weaker than predicted with models, given the fits to the optical Swan bands, with the C\,\textsc{i} 2478\,\AA\ line being especially problematic. The best fit to the UV C\,\textsc{i} lines for Procyon\,B was a factor of $\approx$\,30 lower than the abundance from the optical Swan band fit \citep{Provencal2002}. The fact that UV carbon features are stronger in models than in spectroscopic data seems to be universal for DQs \citep{Dufour2011,Coutu2019,Koester2019}. While this discrepancy has historically been attributed to the inadequacy of the classical van der Waals broadening treatment within the impact approximation, we used models with updated unified C\,\textsc{i} line profiles. Because this discrepancy was not improved with the addition of unified profiles, we artificially decreased the \textit{gf} values in the models by a factor of $\approx$\,10. The \textit{gf} parameter corresponds to the oscillator strengths of transitions between two energy levels. Following this adjustment, the predicted line strengths from the models were similar to those observed for C\,\textsc{i} 1657\,\AA, 1930\,\AA\ and 2478\,\AA. 

We adopted a fitting method for DQs that has been used for many decades \citep{Bergeron1997,Dufour2005,Blouin2019c,Coutu2019,Farihi2026}. In brief, the photometry, parallax and spectral features were fitted independently to constrain the three free parameters, \Teff, \logg, and $\log$(C/He). The main difference between our method and previous studies was the inclusion of the STIS spectra in the photometric part of the fit. Our procedure was as follows: we fixed the models to a starting $\log$(C/He), and performed a combined fit to the STIS spectrum plus the optical and IR photometry. We determined \Teff\ and \logg\ from a fit to the STIS spectrum plus photometry. Then, we fixed the \Teff\ and \logg\ from the initial fit, and fitted the carbon lines in the optical spectrum to determine $\log$(C/He). We generated new models with the new $\log$(C/He) and again fitted the spectrophotometric data to determine \Teff\ and \logg. We repeated these steps until convergence, while additionally varying $\log$(H/He) until the lowest reduced $\chi^2$ was reached, in order to determine limits on hydrogen for all objects. We did not attempt to fit the UV carbon lines to determine abundances due to the issues mentioned above with the line strengths of UV carbon.

\subsubsection{DZ fitting}
\label{sec:DZ_fitting}

Fitting of cool, metal-enriched helium-atmosphere white dwarfs (DZ) requires bespoke model atmosphere grids for each object. The models generated for our fits were from the \citet{Koester2010} code. Our models included several features required for the analysis of DZ white dwarfs, namely calculations for unified line profiles of the Ca\,\textsc{ii}~H and K doublet, Ca\,\textsc{i}\,4227\,\AA, Mg\,\textsc{ii}~H and K doublet, Mg\,\textsc{i}\,b triplet, Mg\,\textsc{i}\,2853\,\AA, and Na\,D doublet, as described in \citet{Hollands2017,Hollands2022}. For the DZ spectra, we used the same iterative fitting procedure as for the DQ white dwarfs, explained in Section~\ref{sec:DQ_fitting}. Due to the low \Teff\ of some of these DZ stars, the metal features could not be fitted independently, as abundances of metals were correlated, and changing one metal abundance affected the shape of other metal lines. In the models we included some metals that are not detected in the spectra but are major rock forming elements (such as silicon and nickel). We fixed the abundances of these undetected elements relative to calcium by assuming ratios of bulk Earth composition. The details of the models used for each individual DZ are discussed in Section~\ref{sec:dz_fits}.

\section{Results}

\subsection{H-atmosphere results}
\label{sec:Hatmosphere_results}

The best-fitting parameters for the H-atmosphere 13\,pc white dwarfs, determined using the fitting methods outlined in Section~\ref{sec:models}, are presented in Table~\ref{tab:13pc_all_parameters_DA}. We fitted STIS spectra alongside \textit{Gaia} DR3 $G$, $G_\mathrm{BP}$, $G_\mathrm{RP}$; 2MASS $J, H, K_\mathrm{s}$; and WISE $W1$, $W2$ photometry, where available, in a combined fit. We fitted the data with two different approaches for most white dwarfs: once including the STIS spectrum, labelled as ‘with UV', and once without it, labelled as ‘without UV' in Table~\ref{tab:13pc_all_parameters_DA}. Additional independent parameters determined from fits to the Balmer lines of DA white dwarfs are also presented in Table~\ref{tab:13pc_all_parameters_DA}.

\begin{table*}
    \centering
        \caption{Best-fit parameters for H-atmosphere white dwarfs within 13\,pc of the Sun. A fit with UV refers to combined spectrophotometric parameters for white dwarfs observed with \textit{HST} STIS or COS. Symbols indicate the photometry used for the fit. No symbol indicates \textit{Gaia}, 2MASS and WISE. $*$ indicates \textit{Gaia} and 2MASS. $\ddag$ indicates $BVRI$ and 2MASS photometry. $\S$ indicates STIS G430L was used in the fit. Any metal features were masked for these fits. Uncertainties are purely statistical. Masses were determined with UV, unless UV data was not available or only high-resolution UV spectra were used.}
    \begin{tabular}{lllllllll}
        \hline
         WD name& SpT & \Teff\ (K)& \logg    & \Teff\ (K)&\logg     &Mass (\Msun)&\Teff\ (K)&\logg   \\
          && (with UV)& (with UV)& (without UV)&(without UV) &&(Balmer)&(Balmer)\\
                  \hline
         0009$+$501 & DAH& 6837\,$\pm$\,6& 8.365\,$\pm$\,0.004& 6674\,$\pm$\,42& 8.313\,$\pm$\,0.016 &0.825\,$\pm$\,0.003& 6529\,$\pm$\,62&8.35\,$\pm$\,0.08\\
         0141$-$675 & DAZ& 6705\,$\pm$\,3& 8.145\,$\pm$\,0.003& 6444\,$\pm$\,25& 8.021\,$\pm$\,0.010 &0.680\,$\pm$\,0.002& 6410\,$\pm$\,4&7.859\,$\pm$\,0.006\\
         0245$+$541 & DAZ& 5211\,$\pm$\,4& 8.224\,$\pm$\,0.003& 5115\,$\pm$\,19& 8.167\,$\pm$\,0.013 &0.724\,$\pm$\,0.002& --&--\\
         0310$-$688 & DAZ& 15\,714\,$\pm$\,5& 8.041\,$\pm$\,0.001& 16\,449\,$\pm$\,158& 8.135\,$\pm$\,0.011 &0.696\,$\pm$\,0.007& 16\,205\,$\pm$\,12&8.12\,$\pm$\,0.003\\
         0413$-$077 $*$ &DA& 16\,333\,$\pm$\,6&  7.900\,$\pm$\,0.001& 16\,697\,$\pm$\,339&7.964\,$\pm$\,0.026 &0.596\,$\pm$\,0.015&16\,959\,$\pm$\,10&7.981\,$\pm$\,0.002\\
         0553$+$053 &DAH& 6112\,$\pm$\,3&  8.346\,$\pm$\,0.004& 5827\,$\pm$\,32&8.213\,$\pm$\,0.014 &0.811\,$\pm$\,0.003&--&--\\
         0642$-$166 $\S$ & DA& 25\,199\,$\pm$\,4& 8.60\,$\pm$\,0.3& 25\,527\,$\pm$\,33& 8.640\,$\pm$\,0.016 &1.022\,$\pm$\,0.009& 25\,762\,$\pm$\,106&8.723\,$\pm$\,0.025\\
         0727$+$482\,A $\ddag$ & DA& 5015\,$\pm$\,17& 7.839\,$\pm$\,0.064& 4966\,$\pm$\,87& 7.941\,$\pm$\,0.061 &0.544\,$\pm$\,0.036& --&--\\
         0727$+$482\,B $\ddag$ & DA& 5064\,$\pm$\,70& 8.17\,$\pm$\,0.14& 4979\,$\pm$\,106& 8.162\,$\pm$\,0.067 &0.682\,$\pm$\,0.044& --&--\\
         0752$-$676 &DA& 5754\,$\pm$\,1&  8.034\,$\pm$\,0.002& 5687\,$\pm$\,10&8.001\,$\pm$\,0.006 &0.607\,$\pm$\,0.001&5634\,$\pm$\,199&7.986\,$\pm$\,0.011\\
         0810$-$353& DAH& --& --& 6503\,$\pm$\,43& 8.285\,$\pm$\,0.020 &0.771\,$\pm$\,0.013& --&--\\
         0821$-$669& DA& 5079\,$\pm$\,5& 8.142\,$\pm$\,0.004& 4960\,$\pm$\,26& 8.061\,$\pm$\,0.017 &0.670\,$\pm$\,0.003& --&--\\
         0839$-$327& DA& 9527\,$\pm$\,4& 7.931\,$\pm$\,0.002& 9196\,$\pm$\,36& 7.815\,$\pm$\,0.009 &0.559\,$\pm$\,0.001& 9227\,$\pm$\,9&7.786\,$\pm$\,0.046\\
         1202$-$232 &DAZ& 8973\,$\pm$\,8&  8.128\,$\pm$\,0.003& 8539\,$\pm$\,75&7.972\,$\pm$\,0.022 &0.675\,$\pm$\,0.002&8637\,$\pm$\,4&7.911\,$\pm$\,0.006\\
         1334$+$039& DA& 5101\,$\pm$\,3& 8.035\,$\pm$\,0.003& 5011\,$\pm$\,15& 7.969\,$\pm$\,0.012 &0.602\,$\pm$\,0.002& --&--\\
         1345$+$238 &DC& 4797\,$\pm$\,7&  7.964\,$\pm$\,0.006& 4725\,$\pm$\,39&7.910\,$\pm$\,0.028 &0.557\,$\pm$\,0.004&--&--\\
         1620$-$391 & DAZ& 23\,238$\pm$\,25& 7.935\,$\pm$\,0.003& 25\,011\,$\pm$\,527& 8.071\,$\pm$\,0.032 &0.675\,$\pm$\,0.020& 24\,651\,$\pm$\,26&8.2\,$\pm$\,0.1\\
         1630$+$089& DA& --& --& 5664\,$\pm$\,9& 8.075\,$\pm$\,0.006 &0.632\,$\pm$\,0.004& 5571\,$\pm$\,76&7.926\,$\pm$\,0.016\\
         1647$+$591& DAV& 12\,408\,$\pm$\,17& 8.32\,$\pm$\,0.02& 12\,240\,$\pm$\,107& 8.304\,$\pm$\,0.012 &0.796\,$\pm$\,0.008& 12\,283\,$\pm$\,35&8.324\,$\pm$\,0.009\\
         1953$-$011 & DAH& 8382\,$\pm$\,8& 8.400\,$\pm$\,0.004& 8000\,$\pm$\,69& 8.242\,$\pm$\,0.020 &0.852\,$\pm$\,0.003& 7688\,$\pm$\,28&8.2\,$\pm$\,0.1\\
         2150$+$591 $*$ & DAH& --& --& 5244\,$\pm$\,33& 8.103\,$\pm$\,0.021 &0.646\,$\pm$\,0.014& --&--\\
         2359$-$434 & DAH& 8941\,$\pm$\,7& 8.530\,$\pm$\,0.003& 8683\,$\pm$\,70& 8.429\,$\pm$\,0.016 &0.937\,$\pm$\,0.002& 8383\,$\pm$\,5&8.265\,$\pm$\,0.005\\
        \hline
    \end{tabular}
    \label{tab:13pc_all_parameters_DA}
\end{table*}

\subsubsection{Combined fits}

\begin{figure*}
    \centering
	\includegraphics[width=\textwidth]{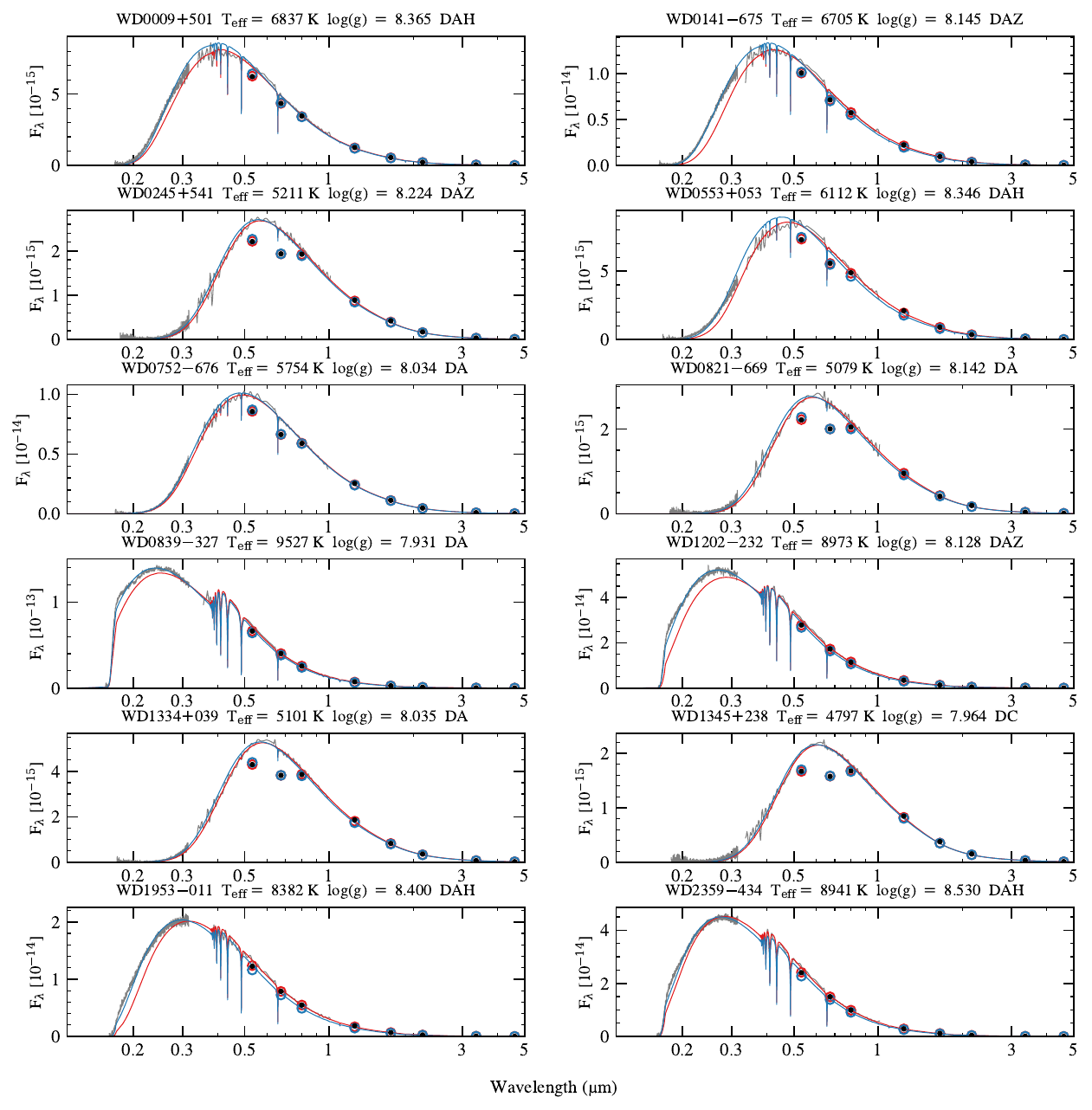}
	\caption{Combined spectrophotometric fits of H-atmosphere white dwarfs in the 13\,pc sample with flux-calibrated STIS G230L spectra. Flux is given in units of erg\,cm$^{-2}$\,s$^{-1}$\,\AA$^{-1}$. The blue models are best fits to STIS spectra plus \textit{Gaia}, 2MASS, and WISE photometry plus parallaxes, where available. The red models are the fits to just the \textit{Gaia}, 2MASS, and WISE photometry plus parallaxes. The blue and red circles show how well these same models fit to the photometry (black dots). All parameters in this figure are from the fits to STIS plus photometry (blue models). \gaia\ XP spectra are shown for completeness but are not incorporated into the fits.}
    \label{fig:plot_all_DA_G230L}
\end{figure*}

\begin{figure}
    \centering
	\includegraphics[width=\columnwidth]{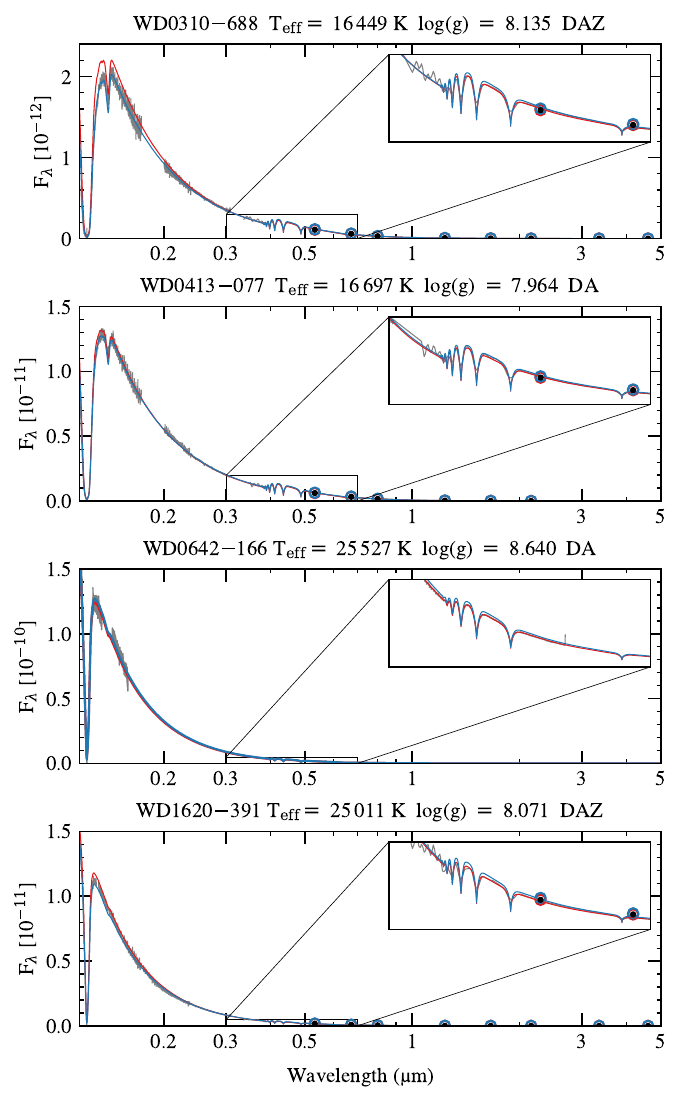}
	\caption{Combined spectrophotometric fits of H-atmosphere white dwarfs in the 13\,pc sample with high-resolution STIS spectra. Flux is given in units of erg\,cm$^{-2}$\,s$^{-1}$\,\AA$^{-1}$. The blue models are best fits to STIS spectra plus \textit{Gaia}, 2MASS, and WISE photometry, where available. The red models are the fits to just the \textit{Gaia}, 2MASS, and WISE photometry. The blue and red circles show how well these same models fit to the photometry (black dots). All parameters in this figure are from the fits to just optical/IR data (red models), as these do not suffer from issues of flux calibration. For WD\,0642$-$166, the red model is the fit to just the flux-calibrated G430L spectrum, and the blue model is the fit to the G430L spectrum plus the E140H spectrum that has been scaled.}
    \label{fig:plot_all_DA_echelle}
\end{figure}

Figures~\ref{fig:plot_all_DA_G230L} and \ref{fig:plot_all_DA_echelle} show the best-fit models for two different fits of H-atmosphere white dwarfs: one including STIS spectra plus optical and IR photometry (blue models), and another for just optical and IR photometry (red models). Low-resolution \textit{Gaia} XP spectra are also displayed in Figs.~\ref{fig:plot_all_DA_G230L} and \ref{fig:plot_all_DA_echelle}. \textit{Gaia} XP spectra provide a visual insight into the performance of the models, but their flux calibration has not been validated as extensively across data releases and external surveys (see, e.g. \citealt{Elms2024}) as \textit{Gaia} photometry. Since the photometry itself is partly derived from XP spectra, we chose to fit the \textit{Gaia} photometry instead and consequently the XP spectra were not incorporated into the combined fit. For plotting the XP spectra, we applied corrections derived using CALSPEC and LAMOST data \citep{Huang2024}. In all plots, the XP spectra were cropped at the blue end to remove noise. Figure~\ref{fig:plot_all_DA_G230L} shows fits to the flux-calibrated, low-resolution STIS G230L spectra, and Fig.~\ref{fig:plot_all_DA_echelle} shows fits to the high-resolution STIS E140M spectra. For some of the coolest spectra, the fits with and without STIS show significant overlap in Fig~\ref{fig:plot_all_DA_G230L}, as the STIS flux is minimal and does not significantly affect the overall fit to the SED. 

For WD\,0642$-$166 (Sirius\,B), all photometry was at least partially contaminated, so the fits presented in Fig.~\ref{fig:plot_all_DA_echelle} are to the G430L spectrum plus the suboptimally flux-calibrated STIS E140H spectrum scaled to the flux-calibrated G430L spectrum (blue), and the G430L spectrum independently (red). From the STIS G430L fit, we calculated a mass of 1.022\,\Msun\ for Sirius\,B, and we also determined a mass of 1.069\,\Msun\ from fits to Balmer lines. An astrometric mass of 1.018\,$\pm$\,0.011\,\Msun\ was determined for Sirius\,B using \textit{HST}, which is independent from atmosphere models \citep{Bond2017b}. An astrometric mass was determined for WD\,0413$-$077 (40\,Eri\,B) of 0.573\,$\pm$\,0.018\,\Msun\ \citep{Bond2017}, which compares to our values of 0.560\,\Msun\ (with UV), 0.596\,\Msun\ (without UV), and 0.606\,\Msun\ (Balmer fit).

\begin{figure}
    \centering
	\includegraphics[width=\columnwidth]{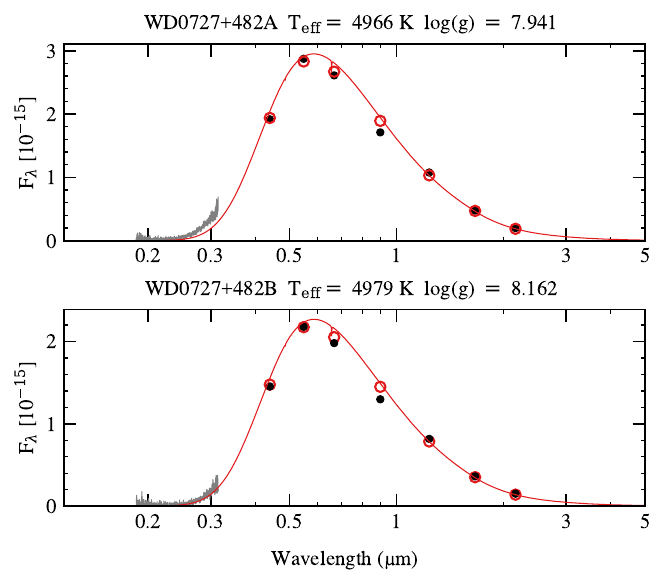}
	\caption{Photometric fits for the wide DA + DA system WD\,0727$+$482 A+B. Reliable \textit{Gaia} photometry is not available for these stars so archival $BVRI$ optical photometry was used in the fit instead. The red circles show how well these same models fit to the photometry (black dots). The STIS spectra were not included in the fit but are plotted for completeness. Flux is given in units of erg\,cm$^{-2}$\,s$^{-1}$\,\AA$^{-1}$.}
    \label{fig:plot_0727}
\end{figure}

Photometric fits to the components of the resolved binary WD\,0727$+$482\,A+B are shown in Fig.~\ref{fig:plot_0727}. These two white dwarfs are part of a quadruple star system with a pair of M dwarfs, and we took the \gaia\ parallax of these M dwarfs to be equivalent to those of WD\,0727$+$482\,A+B in Table~\ref{tab:13pc_info}, as the white dwarfs did not have \gaia\ parallaxes. The white dwarfs also had no \gaia\ photometry, but were observed in $BVRI$ bands \citep{Bergeron2001}, which we incorporated into the fits. 

\begin{figure}
    \centering
	\includegraphics[width=\columnwidth]{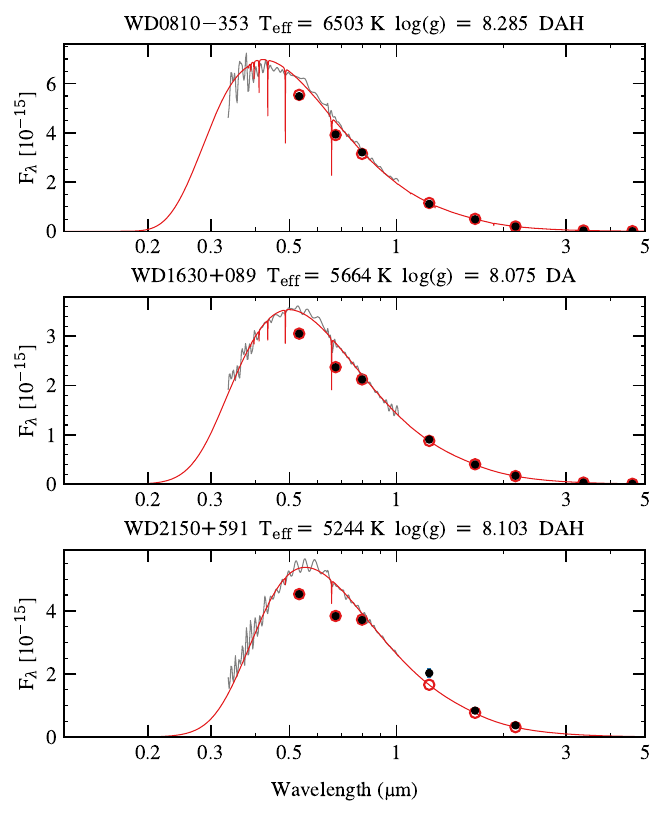}
	\caption{Fits to photometry of DA white dwarfs within 13\,pc for which no STIS spectra are available. Flux is given in units of erg\,cm$^{-2}$\,s$^{-1}$\,\AA$^{-1}$. The red models show the overall fits, and the red circles show how well the models fit to the photometry (black dots).}
    \label{fig:plot_DA_no_STIS_available}
\end{figure}

\begin{figure}
    \centering
	\includegraphics[width=\columnwidth]{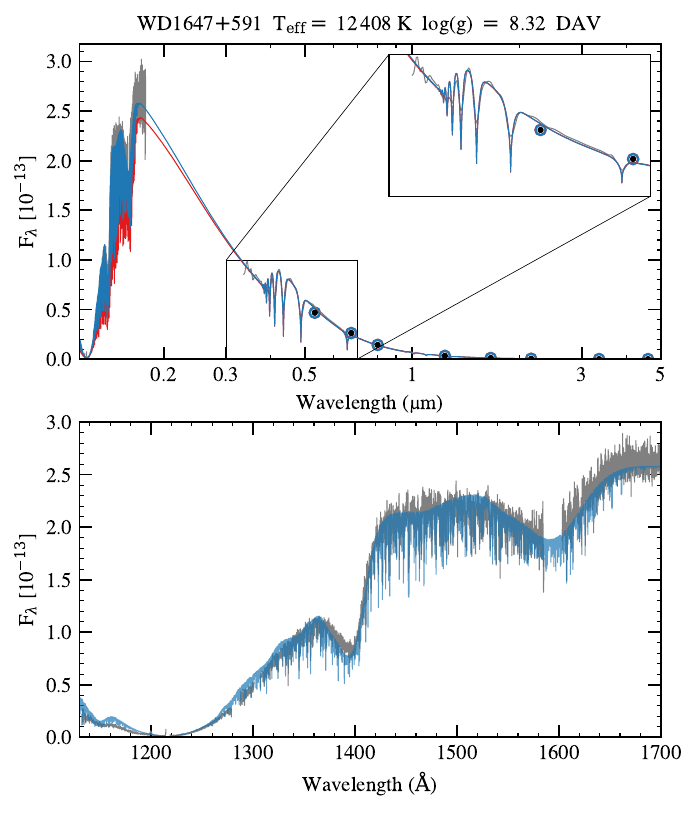}
	\caption{Fit to the COS G130M and G160M spectra plus photometry of WD\,1647$+$591, using \citet{Tremblay2013} models with additional H\textsubscript{2} molecular lines. The blue model is the best fit to the COS spectrum. The red model is the fit to just the \textit{Gaia}, 2MASS, and WISE photometry. The blue and red circles show how well these same models fit to the photometry (black dots). The parameters in this figure are from the fit to the COS spectrum (blue models). The upper panel shows the full SED fit, and the lower panel is a zoom-in of the region with H\textsubscript{2} molecular lines. Flux is given in units of erg\,cm$^{-2}$\,s$^{-1}$\,\AA$^{-1}$}
    \label{fig:plot_DAV_h2}
\end{figure}

Three 13\,pc H-atmosphere white dwarfs, WD\,0810$-$353, WD\,1630$+$089 and WD\,2150$+$591, did not have STIS spectra available, so we fitted just their photometry in Fig.~\ref{fig:plot_DA_no_STIS_available}. WD\,2150$+$591 shows a 2MASS $J$-magnitude excess above the model in Fig.~\ref{fig:plot_DA_no_STIS_available}, which we believe was caused by contamination from a nearby bright star. The DAV WD\,1647$+$591 was not observed with STIS so we instead applied a combined fitting method with the COS spectrum and photometry in Fig.~\ref{fig:plot_DAV_h2}. This white dwarf shows many molecular H$_2$ lines in its high-resolution COS spectrum that were implemented in our models.

\subsubsection{Fits to Balmer lines}

\begin{figure*}
    \centering
	\includegraphics[width=\textwidth]{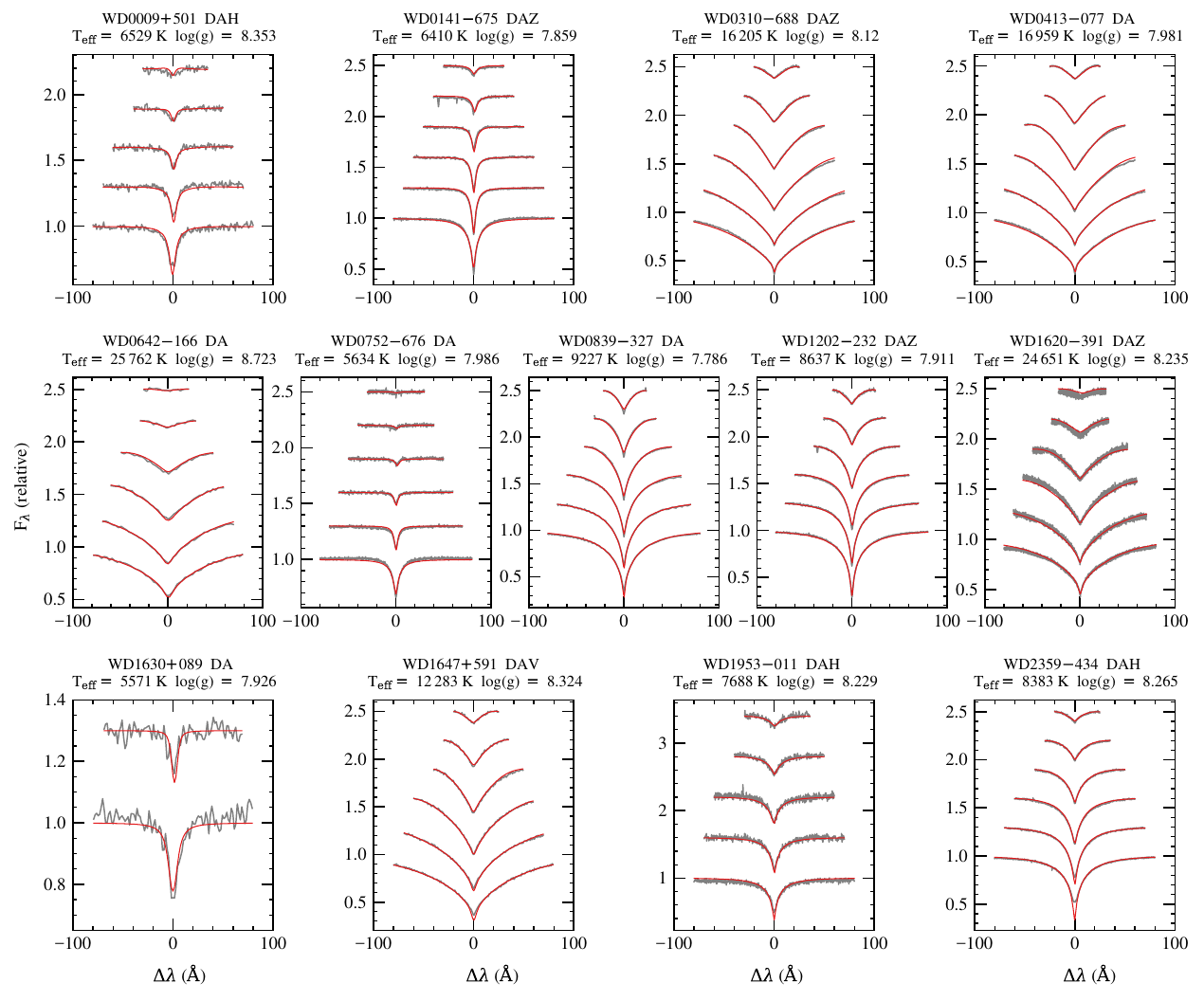}
	\caption{Fits of Balmer lines for white dwarfs in the 13\,pc sample. Flux is given in units of erg\,cm$^{-2}$\,s$^{-1}$\,\AA$^{-1}$. Balmer lines are offset vertically for visual clarity, with H\,$\alpha$ at the bottom of each plot.}
    \label{fig:plot_all_DA_balmer_fits}
\end{figure*}

Fits to the Balmer lines of DA white dwarfs from optical spectroscopy are shown in Fig.~\ref{fig:plot_all_DA_balmer_fits}. The wings of the lines fitted well in all cases, but the cores of the lines in some stars were not well-modelled, possibly due to neglecting non-LTE effects and 3D overshoot \citep{Tremblay2013}. However, this generally had a small effect on the resulting best fit parameters, with possible differences also due to data flux calibration or Zeeman splitting of the line cores. These fits provide an extra set of parameters for single DA white dwarfs in Table~\ref{tab:13pc_all_parameters_DA}. There are known discrepancies between photometric and spectroscopic fits of DA white dwarfs, see \citet{OBrien2023} for details. 

\begin{table*}
    \centering
        \caption{Parameters from combined spectroscopic and photometric fits to the double lined system WD\,0135$-$052. UVES spectra of the two epochs were fitted individually.}
    \begin{tabular}{lllll}
        \hline
          Observation&\Teff$_{,1}$\ (K)&\logg$_{1}$ & \Teff$_{,2}$\ (K)&\logg$_{2}$\\
         HJD [days]&(Balmer)&(Balmer) & (Balmer)&(Balmer) \\
                  \hline
          51737.8611& 7504$^{+126}_{-125}$&8.056$^{+0.040}_{-0.083}$& 7243$^{+51}_{-87}$&7.701$^{+0.030}_{-0.026}$\\
          51741.8696& 7403$^{+72}_{-57}$&7.755$^{+0.025}_{-0.026}$& 7080$^{+88}_{-170}$&7.917$^{+0.036}_{-0.035}$\\
         \hline
    \end{tabular}
    \label{tab:13pc_all_parameters_DWD}
\end{table*}

\begin{figure*}
    \centering
	\includegraphics[width=\textwidth]{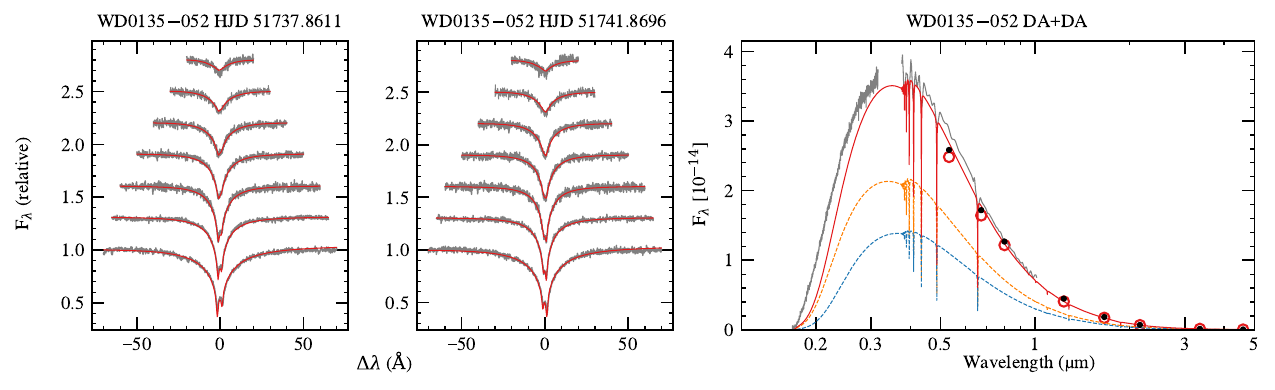}
	\caption{Combined fits of Balmer lines and photometry for the unresolved double-lined double white dwarf WD\,0135$-$052 in the 13\,pc sample. Flux is given in units of erg\,cm$^{-2}$\,s$^{-1}$\,\AA$^{-1}$. Balmer lines are offset vertically for visual clarity, with H\,$\alpha$ at the bottom of each plot. The STIS spectrum was not included in the fit but is shown on the plot. The orange and blue dashed models are the two individual models that best fit to the spectrum at HJD\,51741.8696, which are then combined to form the red model. The red circles show how well the overall model fits to the photometry (black dots).}
    \label{fig:plot_all_DWD_balmer_fits}
\end{figure*}

In Table~\ref{tab:13pc_all_parameters_DWD}, we provide the best-fit parameters for both components of the well-known double-lined double degenerate system WD\,0135$-$052 \citep{Saffer1988}, from the fit to each of the two UVES spectra plus photometry. We fitted this system using the \textsc{wd$-$bass} code \citep{Munday2024}. The fits to the Balmer lines of WD\,0135$-$052 are shown in Fig.~\ref{fig:plot_all_DWD_balmer_fits}, where we also show the two models that combined to fit the spectrum observed at epoch HJD\,51741.8696, compared to the full SED of WD\,0135$-$052. The error bars on the parameters are large due to the two white dwarfs having a similar \Teff, which introduced degeneracies into the fit.

\begin{table*}
    \centering
        \caption{Photospheric metal abundances of hydrogen-atmosphere white dwarfs within 13\,pc of the Sun. \Teff\ and \logg\ for these stars are listed in Table~\ref{tab:13pc_all_parameters_DA}.}
    \begin{tabular}{lllllll}
        \hline
         WD name&  SpT & $\log$(Ca/H)& $\log$(Mg/H)&$\log$(O/H)&$\log$(Si/H)&$\log$(C/H)\\
        \hline
        0141$-$675&  DAZ& $-$10.7\,$\pm$\,0.02 & --& --& -- &-- \\
        0310$-$688& DAZ&    --& $-$7.77\,$\pm$\,0.09 & $-$7.14\,$\pm$\,0.12 & $-$7.93\,$\pm$\,0.06 & --\\
        1202$-$232& DAZ& $-$9.7\,$\pm$\,0.01 & --& --& --&--\\
        1620$-$391&  DAZ&   --& -- &  -- & $-$7.41\,$\pm$\,0.15 & $-$7.73\,$\pm$\,0.08 \\
        \hline
    \end{tabular}
    \label{tab:13pc_all_parameters_DAZ}
\end{table*}

\subsubsection{Fits to photospheric metal lines}

The metal abundances from fits to DAZ white dwarfs are presented in Table~\ref{tab:13pc_all_parameters_DAZ}. WD\,0141$-$675 and WD\,1202$-$232 displayed only calcium lines in the optical and no photospheric lines in the UV. WD\,0310$-$688 and WD\,1620$-$391 appeared to be pure-H atmosphere DAs in the optical, but displayed many photospheric metal lines in the UV. We did not have access to any Ca\,\textsc{ii}~H and K line coverage of the DAZ white dwarf WD\,0245$+$541, however the calcium abundance of this star was measured as $\log$(Ca/H)\,$=$\,$-$12.7 using a Keck HIRES spectrum \citep{Zuckerman2003}.

None of the H-rich DAZ white dwarfs in our sample have metal lines newly-identified from this work. WD\,0141$-$675 has been well-studied, as it has a now-retracted astrometric planet candidate from \gaia\ DR3 \citep{Rogers2024,Ramirez2025}. For WD\,0141$-$675, optical fits under-predicted the STIS spectrum flux \citep{Rogers2024}. The calcium abundance of WD\,1202$-$232 was measured using a Keck HIRES spectrum \citep{Zuckerman2003}. WD\,0310$-$688 was previously reported to show a mid-IR excess from \textit{JWST} MIRI, due to a giant planet candidate or cold debris disk \citep{Limbach2024}; the same STIS spectrum discussed in their work was analysed and fitted here. \citet{Debes2025} analysed the STIS spectrum of WD\,1620$-$391, and also did not detect an infrared excess with \textit{JWST} MIRI.

\begin{figure*}
    \centering
	\includegraphics[width=\textwidth]{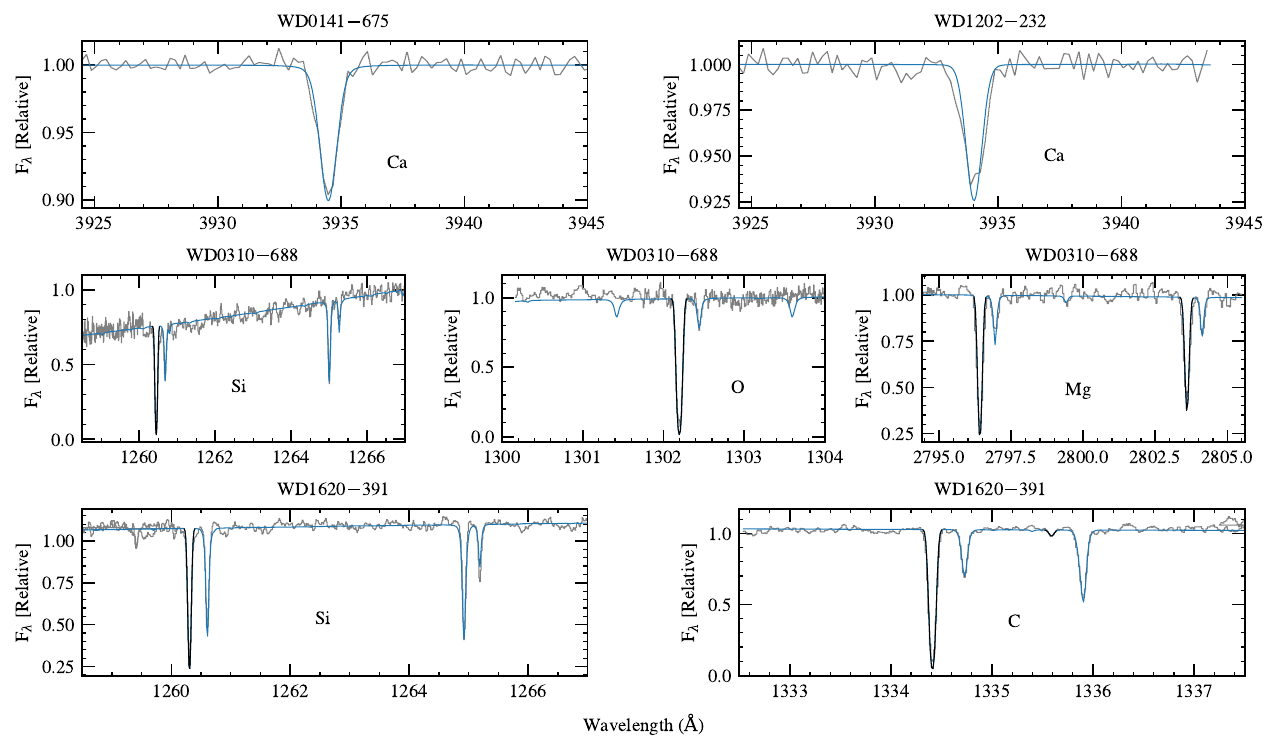}
	\caption{Fits to metal lines in H-atmosphere white dwarfs in the 13\,pc sample. Flux is given in units of erg\,cm$^{-2}$\,s$^{-1}$\,\AA$^{-1}$. The blue models are the fits to the photospheric optical and UV metal features. The black models are Gaussian fits to the interstellar lines.}
    \label{fig:plot_DAZ_metal_line_fits}
\end{figure*}


\subsection{He-atmosphere results}
\label{sec:Heatmosphere_results}

\subsubsection{DC results}
\label{sec:dc_fits}

\begin{figure*}
    \centering
	\includegraphics[width=\textwidth]{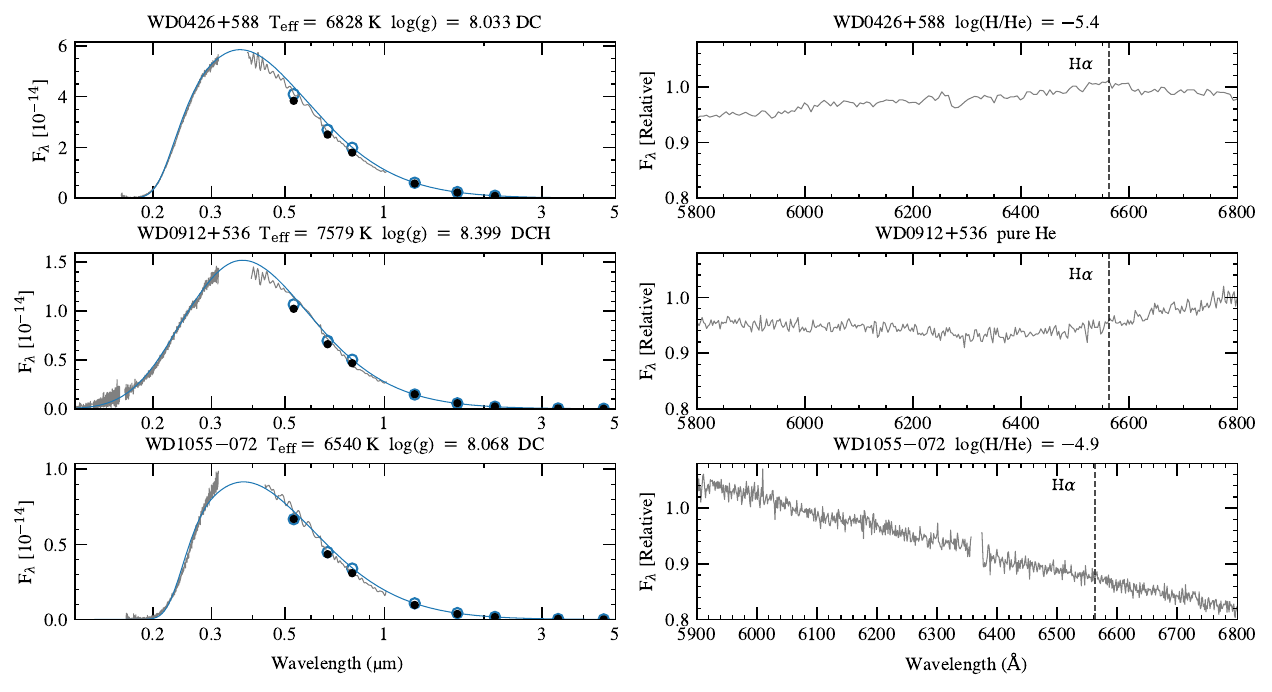}
	\caption{Combined spectrophotometric fits of white dwarfs in the 13\,pc sample with He-rich atmospheres that fit best with He ($+$\,H) models. Flux is given in units of erg\,cm$^{-2}$\,s$^{-1}$\,\AA$^{-1}$. \textit{Left}: combined fits to the full SED, where the blue models are best fits to STIS spectra plus \textit{Gaia}, 2MASS, and WISE photometry, where available. The blue circles show how well these same models fit to the photometry (black dots). \textit{Right}: optical spectra covering the H\,$\alpha$ region, where the dashed lines indicate where an H\,$\alpha$ feature would be present. The 100\,MG field in WD\,0912$+$536 would distort and split H\,$\alpha$.}
    \label{fig:plot_DC}
\end{figure*}

Figure~\ref{fig:plot_DC} shows the best fits to the DC white dwarfs with He-rich atmospheres in 13\,pc. Table~\ref{tab:13pc_all_parameters_DC} lists the best-fitting parameters for the DC white dwarfs. We found that the best-fitting models to just the optical and IR photometry did not reliably predict the UV flux, and therefore we did not include parameters for fits without STIS in Table~\ref{tab:13pc_all_parameters_DC}. The H\,$+$\,He DC models are sensitive to the Ly\,$\alpha$ opacity, especially in the UV, which makes them responsive to the presence of trace hydrogen, and therefore adding the STIS spectra to our fits enabled us to constrain the hydrogen content.

WD\,0426$+$588 and WD\,1055$-$072 required hydrogen for a good fit. WD\,0426$+$588 has a model-independent mass measurement of 0.675\,$\pm$\,0.051\,\Msun\ determined from microlensing \citep{Sahu2017}. Our mass from UV fitting is 0.594\,\Msun, which is within 2\,$\sigma$ of the microlensing mass. The magnetic DCH WD\,0912$+$536, with a 100\,MG field strength \citep{Angel1972,Bagnulo2020}, fitted well with pure-He models. The magnetic field was identified with continuum polarisation measurements, and the strong field does not alter the shape of the SED due to a lack of spectral features. 

We tested the effects of adding carbon rather than hydrogen to the DC models, as trace carbon is known to reproduce UV cooling tracks \citep{Blouin2023b,Camisassa2023}. Helium and trace carbon models were tested on the two DCs that fitted well with $\log$(H/He)\,$\approx$\,$-$5, WD\,0426$+$588 and WD\,1055$-$072, and they produced substantially worse fits. We then generated C $+$ H $+$ He models with \Teff, \logg, and $\log$(H/He) fixed at their best-fitting values from the fits with H $+$ He models, and $\log$(C/He) as the only free parameter. Fitting with these models provided an upper limit for the carbon content: $\log$(C/He)\,$=$\,$-$8.4 for WD\,0426$+$588, and $\log$(C/He)\,$=$\,$-$9.0 for WD\,1055$-$072. These limits are discussed in the context of DQs in Section~\ref{sec:carbon_hydrogen}. 


\begin{table*}
    \centering
        \caption{Best-fit parameters for He-rich DC white dwarfs within 13\,pc of the Sun. The \Teff\ and \logg\ are combined spectrophotometric parameters determined from fitting STIS spectra plus optical and IR photometry. Symbols indicate the photometry used for the fit. No symbol indicates \textit{Gaia}, 2MASS and WISE. $*$ indicates \textit{Gaia} and 2MASS/JHK photometry.}
    \begin{tabular}{llllll}
        \hline
         WD name&  SpT & \Teff\ (K)& \logg &Mass (\Msun)&$\log$(H/He)\\
        \hline
        0426$+$588 $*$ &  DC& 6828\,$\pm$\,2&  8.033\,$\pm$\,0.002 & 0.594\,$\pm$\,0.001 &$-$5.4\,$\pm$\,0.5\\
        0912$+$536& DCH& 7579\,$\pm$\,3&   8.399\,$\pm$\,0.002 & 0.834\,$\pm$\,0.001 &pure He\\
        1055$-$072&  DC& 6540\,$\pm$\,3&  8.068\,$\pm$\,0.004 & 0.615\,$\pm$\,0.002 &$-$4.9\,$\pm$\,0.5\\
        \hline
    \end{tabular}
    \label{tab:13pc_all_parameters_DC}
\end{table*}

\subsubsection{DQ results}
\label{sec:dq_fits}

\begin{figure*}
    \centering
	\includegraphics[width=\textwidth]{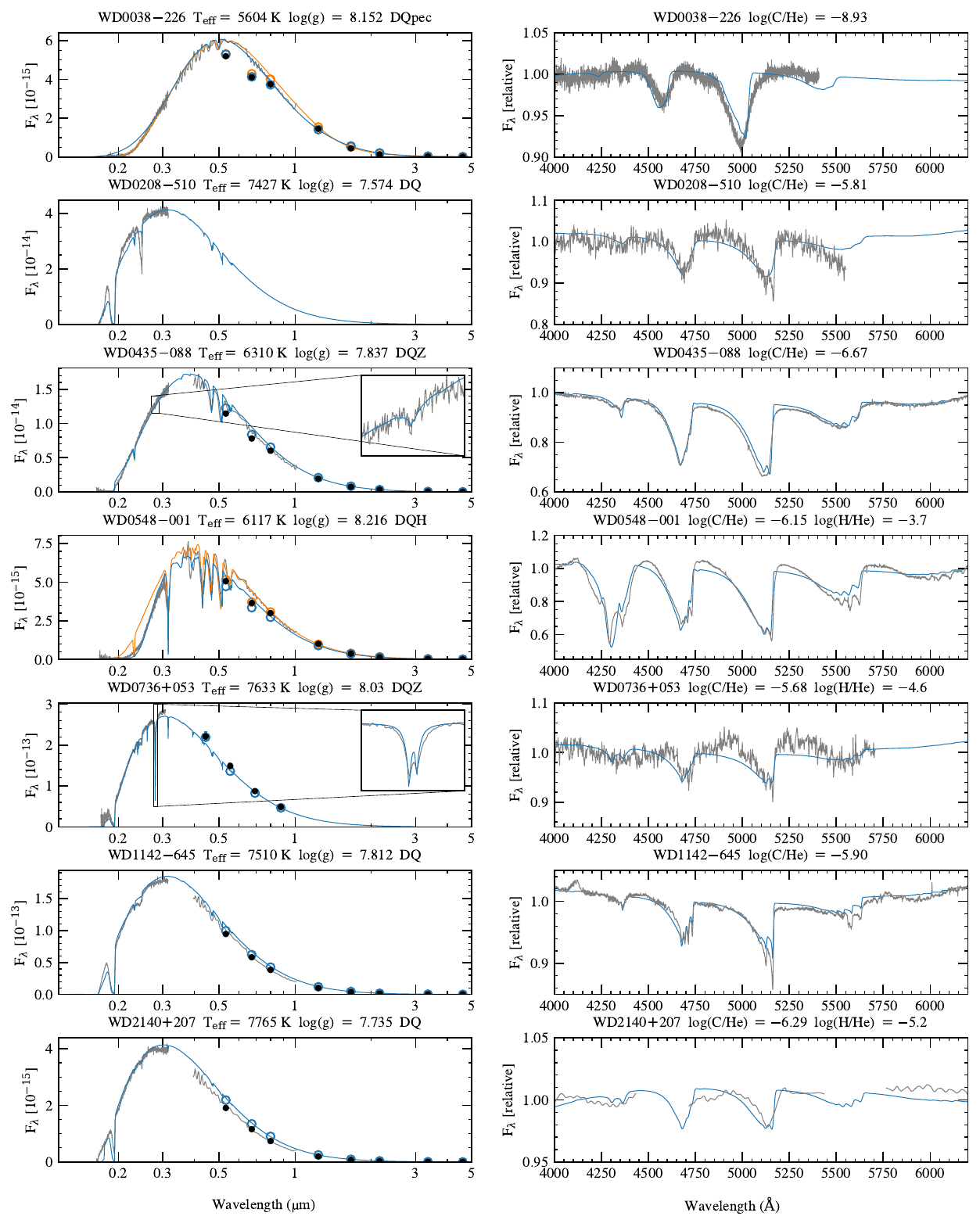}
	\caption{Combined spectrophotometric fits of DQ white dwarfs in the 13\,pc sample with carbon absorption lines. \textit{Left}: combined fits to the full SED. \textit{Right}: fits to the optical Swan bands. Inset plots show the Mg\,\textsc{ii}\,2800\,\AA\ doublet. The carbon lines in the UV have been adjusted to match the observations. Flux is given in units of erg\,cm$^{-2}$\,s$^{-1}$\,\AA$^{-1}$. The blue models show our fits to the optical and UV data, and the orange models are independent fits to two of the same stars from \citet{Farihi2026}. The blue and orange circles show how well these same models fit to the photometry (black dots).}
    \label{fig:plot_DQ}
\end{figure*}

The best-fitting parameters and abundances for the DQ white dwarfs within 13\,pc are presented in Table~\ref{tab:13pc_all_parameters_DQ} and fits are shown in Fig.~\ref{fig:plot_DQ}. Each DQ white dwarf required tailored modelling, and we discuss them individually below.

\textbf{WD\,0038$-$226}: This star has the weakest carbon abundance of any known DQ white dwarf. A previous fit to an optical spectrum of WD\,0038$-$226 provided a $\log$(C/He) of $-$8.4, which is 0.5\,dex higher than the abundance determined in this work \citep{Blouin2019c}. This discrepancy is possibly due to the differences in the models, as WD\,0038$-$226 displays CIA \citep{Giammichele2012}, the effects of which are challenging to model. Our solution of 5600\,K and 0.65\,\Msun\ agrees with an independent mass$-$\Teff\ relation for WD\,0038$-$226 \citep{Blouin2024}. The shape of the UV flux of WD\,0038$-$226 fitted better with trace hydrogen added to the models at $\log$(H/He)\,$=$\,$-$6.5. WD\,0038$-$226 was observed with the \textit{HST}/FOS in the UV, and it was also found that a low hydrogen abundance provided a better fit compared with zero hydrogen \citep{Wolff2002}. However, any amount of hydrogen greatly suppressed the strength of the Swan bands compared to the observations, and none of our models could fit the Swan bands as well as a hydrogen-free model did. Additionally, \textit{JWST} observations showed that for WD\,0038$-$226, $\log$(H/He)\,$>$\,$-$5 would produce a 2.4\,$\mu$m feature which was not detected \citep{Blouin2024}. Therefore, we elected not to include hydrogen in our final fit to WD\,0038$-$226. Recently, a new fit to WD\,0038$-$226, with high resolution optical and near infrared spectra and improved atmosphere models from a detailed analysis of a representative sample of DQ and DQpec stars, indicates that helium CIA influences the shape of the infrared flux suppression \citep{Farihi2026}. This new fit (plotted in orange in Fig.~\ref{fig:plot_DQ}), has a lower but consistent \Teff\ with other studies, and demonstrates that $\log$(H/He)\,$=$\,$-$6.8 matches the UV data, while a carbon abundance of $\log$(C/He)\,$=$\,$-$8.9 (comparable to our value) successfully reproduces the Swan bands \citep{Farihi2026}.

\textbf{WD\,0208$-$510}: This DQ is in a close binary with a K dwarf companion, making optical spectroscopic observations challenging. A fit to the combination of STIS optical G430L spectra plus UV and optical photometric Wide Field Camera 3 (WFC3) UVIS points gave a \logg\ of 8.02 and $\log$(C/He) of $-$4.8 \citep{Farihi2013}. Our fits to the STIS UV G230L spectra without photometry produced a much lower \logg\ of 7.574, and a lower $\log$(C/He) of $-$5.81. In an attempt to reconcile this difference, we fixed the \logg\ to 8 and fitted for \Teff\ and $\log$(C/He). We were unable to obtain a satisfactory fit to both the UV spectrum and the optical Swan bands. The $\log$(C/He) determined from the fit to the STIS G430L plus WFC3 UVIS data provided a good fit to the overall SED but the Swan bands were much weaker and required a $\log$(C/He) of 1\,dex lower to fit well with our models \citep{Farihi2013}. Therefore, we decided to present the parameters determined in our initial fit, keeping the low \logg\ rather than fixing \logg\ to 8. 

\textbf{WD\,0435$-$088}: We observed a weak Mg\,\textsc{ii} feature at 2800\,\AA\ in the STIS G230L spectrum of this white dwarf, which is shown in an inset plot in Fig.~\ref{fig:plot_DQ}. We therefore recomputed models with low levels of magnesium, with all other parameters kept constant in the models. This is the first detection of photospheric metals in the atmosphere of this DQ white dwarf, adding to the small number of known DQZ white dwarfs alongside WD\,0736$+$053 (Procyon\,B).

\textbf{WD\,0548$-$001}: To fit this DQH white dwarf with CH molecular bands, we generated a small grid covering \Teff\,$=$\,5500\,$-$\,6500\,K in steps of 250\,K, \logg\,$=$\,7.75\,$-$\,8.50 in steps of 0.25 dex, $\log$(C/He)\,$=$\,$-$5.5 to $-$6.5 in steps of 0.25 dex, and $\log$(H/He)\,$=$\,$-$5.5 to $-$6.5 in steps of 0.25 dex. This white dwarf has a magnetic field strength of 7\,MG \citep{Bagnulo2022}, which affects the shape of the molecular features in the optical and UV spectra. We fitted the spectra of WD\,0548$-$001 with non-magnetic models and achieved an acceptable fit. However, the fit to the CH molecular feature near 4300\,\AA\ was sub-optimal. An independent fit to the optical spectrum of this star required substantially less hydrogen for a good fit, $-$5 compared to the $-$3.5 determined in our UV plus optical fit \citep{Farihi2026}. This fit is shown in orange in Fig.~\ref{fig:plot_DQ}.

\textbf{WD\,0736$+$053}: To fit WD\,0736$+$053 (Procyon\,B), we used a grid with fixed \logg\ and metal abundances (magnesium, iron, calcium), with \Teff, $\log$(C/He) and $\log$(H/He) allowed to vary. We chose to fix some parameters in the models due to the large number of free parameters. The DQZ grid spanned \Teff\,$=$\,7250\,$-$\,8250\,K in steps of 250\,K, $\log$(C/He)\,$=$\,$-$5.0 to $-$6.0 in steps of 0.25 dex, and $\log$(H/He)\,$=$\,$-$4.0 to $-$5.0 in steps of 0.5 dex. The \logg\ in the models was fixed at 8.03, derived from a dynamical mass of Procyon\,B \citep{Bond2015}. The metal abundances in the grid were fixed at: $\log$(Mg/He)\,$=$\,$-$10.5, $\log$(Fe/He)\,$=$\,$-$10.7, $\log$(Ca/He)\,$=$\,$-$12.0 \citep{Provencal2002}. We also used $BVRI$ photometry from \textit{HST} Wide Field and Planetary Camera 2 \citep{Provencal1997} in the combined fit, as no \gaia\ photometry is available for this object due to its bright companion. This white dwarf was observed with the STIS CCD G230LB grating because Procyon\,A would saturate a G230L NUV-MAMA observation. The G230LB grating has a known red light scattering issue \citep{Worthey2022}, and is less well flux-calibrated than the G230L grating. Therefore, the red end of the G230LB spectrum is not expected to fit well to the models. A previous reported hydrogen abundance of Procyon\,B, $\log$(H/He)\,$<$\,$-$4, is consistent with our abundance of $\log$(H/He)\,$=$\,$-$4.6 \citep{Provencal2002}.

\textbf{WD\,2140$+$207}: The UVES spectrum of WD\,2140$+$207 suffered from echelle order `ripples' which could be due to issues with the reduction or blaze function. We mitigated for this by binning the spectrum and smoothing it with a Savitzky–Golay filter. As shown in Fig.~\ref{fig:plot_DQ}, there are gaps in the UVES spectrum at the location of two of the Swan bands, centred at 4737\,\AA\ and 5636\,\AA, and therefore only one Swan band could be fitted, which led to degeneracies in the solution. We found that adding hydrogen at an abundance of $\log$(H/He)\,$=$\,$-$5.2 to the C\,$+$\,He models resulted in a better fit to the spectrum and SED of WD\,2140$+$207, and therefore the final best-fitting solution for this white dwarf contains a trace hydrogen abundance. 

Three of the seven DQ white dwarfs in the 13\,pc sample required some level of hydrogen to fit the UV and optical data. This result is in contrast with a larger optical-only study, which found evidence of hydrogen in two out of a sample of 293 optical observations of DQs below 10\,000\,K \citep{Coutu2019}. The CH molecular band near 4300\,\AA\ should be visible for DQs below 8000\,K if $-$4.0\,$<$\,$\log$(H/He)\,$<$\,$-$2.0 \citep{Coutu2019}. In our analysis including UV spectra, WD\,0736$+$053 and WD\,2140$+$207 had $\log$(H/He) values of $-$4.6 and $-$5.2 respectively, which is at a low enough level that the CH feature is not present \citep{Farihi2026}. WD\,0548$-$001 showed a clear CH feature and therefore required a much higher $\log$(H/He) of $-$3.7. 

To test the impact of hydrogen, we fitted the remaining DQ white dwarfs with C $+$ H $+$ He models with \Teff, \logg, and $\log$(C/He) fixed at their best-fitting values from the fits with C $+$ He models, and the hydrogen abundance was allowed to vary, to find limits on the hydrogen content. For WD\,0208$-$510, WD\,0435$-$088, and WD\,1142$-$645, $\log$(H/He)\,$=$\,$-$5.5 was determined as the upper limit of hydrogen, which is consistent with the low hydrogen abundances from \citet{Farihi2026}. 

\begin{figure}
    \centering
	\includegraphics[width=\columnwidth]{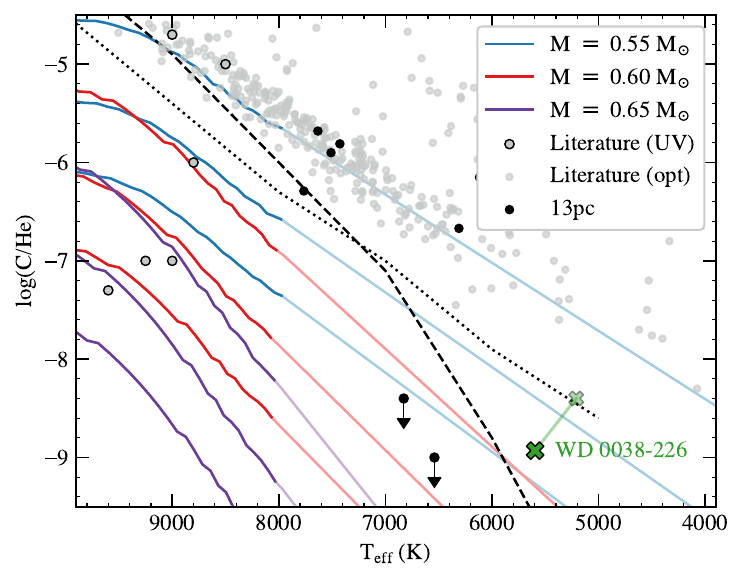}
	\caption{Atmospheric carbon abundance as a function of temperature for DQ white dwarfs. The background sources shown in grey are from \citet{Blouin2019c,Coutu2019,Tremblay2020,OBrien2023,OBrien2024}. The filled black circles represent 13\,pc carbon detections and the arrows represent upper limits on carbon found in the 13\,pc DC white dwarfs. The green cross is our fit to the DQpec WD\,0038$-$226, with the faded cross being a fit to the same object by \citet{Blouin2019c}. The hollow black circles represent DQs with UV carbon detections from the International Ultraviolet Explorer \citep{Weidemann1995}. The solid lines show predicted sequences of carbon dredge-up assuming different white dwarf masses and envelope carbon mass fractions (from bottom to top: 0.2, 0.4, and 0.6; \citealt{Bedard2022_iii, Kilic2025_100pc}). A standard envelope thickness of $\log q =\,-2$ was adopted. The sequences were linearly extrapolated below 8000\,K, where they are shown in a faded colour. The black lines show the carbon detection limits: \citet{Koester2010} (dashed), and \citet{Blouin2018b} (dotted) with SNR\,$=$\,50.}
    \label{fig:carbon_abundances_and_limits}
\end{figure}

Figure~\ref{fig:carbon_abundances_and_limits} shows the carbon abundances in DQ white dwarfs as a function of \Teff. We show the parameters for WD\,0038$-$226 with green crosses, both from this work (using the \citet{Koester2010} code) alongside those from \citet{Blouin2019c}, joined together with a line. This difference indicates the disparity between the two models at low \Teff\ and low carbon abundance. The difference between the two models for various carbon abundances is also shown on the plot, and at lower \Teff\ there are significant deviations.

Excluding the DQpec, four out of the six DQ white dwarfs in this sample have very low masses when fitted with UV data, shown in Table~\ref{tab:13pc_all_parameters_DQ}: WD\,0208$-$510 (0.357\,\Msun), WD\,0435$-$088 (0.480\,\Msun), WD\,1142$-$645 (0.470\,\Msun), and WD\,2140$+$207 (0.431\,\Msun). DQ white dwarfs are expected to have very slightly lower than average masses due to the mass-dependence of carbon dredge up \citep{Bedard2024}. However, optical-only fits do not produce such low masses as those determined in this work \citep{Coutu2019,Farihi2026} . A mass of 0.59\,\Msun\ was previously derived for WD\,0208$-$510 when near-UV fluxes were omitted from the fit \citep{Farihi2013}. Masses of 0.55\,\Msun\ for WD\,0435$-$088, 0.58\,\Msun\ for WD\,1142$-$645, and 0.50\,\Msun\ for WD\,2140$+$207, were determined from fits to optical data alone \citep{Coutu2019}. The analysis of \citet{Farihi2026}, although on a small sample of stars, does not show a mass deviation from that expected for an average white dwarf. The gravitational microlensing technique with \textit{HST} WFC3 observations, a method entirely independent of model atmospheres and spectroscopy, was used to determine a mass of 0.56\,$\pm$\,0.08\,\Msun\ for WD\,1142$-$645 \citep{McGill2023}. The mass of WD\,1142$-$645 that we determined from fitting STIS UV data is 2\,$\sigma$ away from the microlensing mass \citep{McGill2023} and the optical mass \citep{Coutu2019}. 

Our results indicate that including STIS UV data in DQ fitting might bias the solutions towards low \Teff\ and masses. This suggests that there are issues with DQ models in the UV, which can be somewhat mitigated for by adding a substantial amount of hydrogen to the models, hence the discrepancy between our results and \citet{Coutu2019}. Uncertainty in helium ionisation equilibrium could also contribute to this discrepancy \citep{Farihi2026}. As discussed in Section~\ref{sec:DQ_fitting}, we artificially decreased the \textit{gf} values in our models in order to match the UV C\,\textsc{i} line strengths. However, this does not change the full slope of the STIS spectrum, and therefore the resulting best-fit parameters are still affected. 

\begin{table*}
    \centering
        \caption{Best-fit parameters for helium-atmosphere white dwarfs with carbon bands within 13\,pc of the Sun. Symbols indicate the photometry used for the fit. No symbol indicates \textit{Gaia}, 2MASS and WISE. $\ddag$ indicates $BVRI$ photometry was used in the fit. $\S$ indicates no photometry was used in the fit.}
    \begin{tabular}{llllllllll}
        \hline
         WD name&  SpT &\Teff\ (K)&\logg &Mass (\Msun)& $\log$(C/He)& $\log$(H/He)&$\log$(Mg/He)& $\log$(Ca/He)&$\log$(Fe/He)\\
                  \hline
        0038$-$226&  DQpec&5604\,$\pm$\,8&8.152\,$\pm$\,0.008& 0.666\,$\pm$\,0.005 & $-$8.93\,$\pm$\,0.03&  --  &--   & --  &--   \\
        0208$-$510 $\S$ & DQ&7427\,$\pm$\,27&7.574\,$\pm$\,0.02& 0.357\,$\pm$\,0.008 & $-$5.81\,$\pm$\,0.03&   $<-$5.5&--   & --  &--   \\
        0435$-$088& DQZ& 6310\,$\pm$\,5& 7.837\,$\pm$\,0.005& 0.480\,$\pm$\,0.003 & $-$6.67\,$\pm$\,0.01&$<-$5.5&$-$12.7\,$\pm$\,0.2& --  &--   \\
        0548$-$001& DQH& 6117\,$\pm$\,8& 8.216\,$\pm$\,0.007& 0.709\,$\pm$\,0.005 & $-$6.15\,$\pm$\,0.02& $-$3.7\,$\pm$\,0.1&--   & --  &--   \\
        0736$+$053 $\ddag$ & DQZ& 7633\,$\pm$\,3& 8.03 (fixed) & 0.594\,$\pm$\,0.001 & $-$5.68\,$\pm$\,0.03&$-$4.6\,$\pm$\,0.1&$-$10.4 (fixed) & $-$12.0 (fixed)&$-$10.7 (fixed)\\
        1142$-$645& DQ& 7510\,$\pm$\,6& 7.812\,$\pm$\,0.005& 0.470\,$\pm$\,0.003 & $-$5.90\,$\pm$\,0.01&$<-$5.5&--   & --  &--   \\
        1917$-$077& DBQA& 10\,785\,$\pm$\,2& 8.05 (fixed) & 0.612\,$\pm$\,0.001 & $-$5.27\,$\pm$\,0.02&  $-$5.55\,$\pm$\,0.1&--   & --  &--   \\
        2140$+$207&  DQ&7765\,$\pm$\,10&7.735\,$\pm$\,0.007& 0.431\,$\pm$\,0.003 & $-$6.29\,$\pm$\,0.02&  $-$5.2\,$\pm$\,0.1&--   & --  &--   \\
        \hline
    \end{tabular}
    \label{tab:13pc_all_parameters_DQ}
\end{table*}

\subsubsection{DBQA results}

\begin{figure}
    \centering
	\includegraphics[width=\columnwidth]{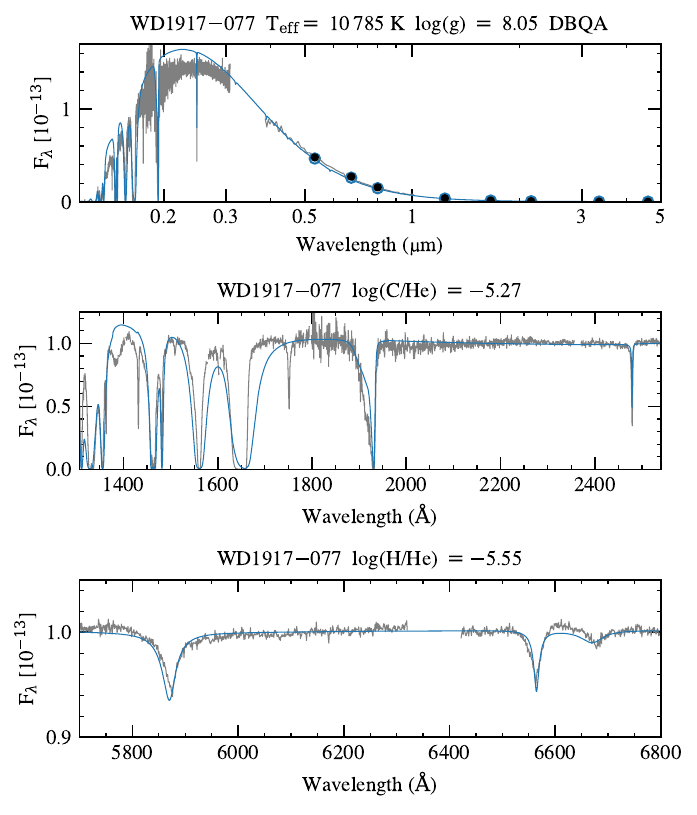}
	\caption{\textit{Top}: A combined spectrophotometric fit to the DBQA white dwarf WD\,1917$-$077. \textit{Middle}: The fit to COS and STIS spectra focusing on the region with carbon features. \textit{Bottom}: The fit to the X-shooter spectrum covering H\,$\alpha$ and the He\,\textsc{ii} 5876\,\AA\ line. In all plots, flux is given in units of erg\,cm$^{-2}$\,s$^{-1}$\,\AA$^{-1}$. The model reproduces the depth of almost all carbon lines, except for C\,\textsc{i} 1432.1\,\AA\ and C\,\textsc{i} 1751.8\,\AA. The blue models show the overall fits, and the blue circles show how well these same models fit to the photometry (black dots).}
    \label{fig:plot_DBQA}
\end{figure}

The DBQA white dwarf WD\,1917$-$077 displays hydrogen and helium features in its optical spectrum and carbon features in its UV spectrum. We generated a grid to fit this white dwarf with \Teff, $\log$(C/He) and $\log$(H/He) allowed to vary, and \logg\ fixed. The DBQA grid spanned \Teff\,$=$\,10\,000\,$-$\,11\,000\,K in steps of 250\,K, $\log$(C/He)\,$=$\,$-$5.0 to $-$6.0 in steps of 0.25 dex, and $\log$(H/He)\,$=$\,$-$5.0 to $-$6.0 in steps of 0.5 dex. The \logg\ in the models was fixed at 8.05, which was determined from a fit to \gaia\ photometry \citep{OBrien2024}. An iterative fitting method was used: the optical and infrared photometry was fitted for \Teff, and the STIS and COS spectra plus the X-shooter spectrum were fitted for $\log$(C/He) and $\log$(H/He). We did not include the STIS and COS spectra in the photometric fit due to suboptimal flux calibration. The best-fitting parameters for WD\,1917$-$077 are in Table~\ref{tab:13pc_all_parameters_DQ}, and the best-fitting model is shown in Fig.~\ref{fig:plot_DBQA}. Our best-fitting model at $\log$(C/He)\,$=$\,$-$5.27 adequately reproduced the depth of almost all carbon lines in the UV, except for C\,\textsc{i} 1432.1\,\AA\ and C\,\textsc{i} 1751.8\,\AA\ (see middle panel of Fig.~\ref{fig:plot_DBQA}). The results of our fit differ to those determined from a fit to solely UV spectra WD\,1917$-$077, where $\log$(C/He)\,$=\,-$6.7\,$\pm$\,0.2 using only one line: C\,\textsc{i} 2479\,\AA, observed with the International Ultraviolet Explorer spectrograph \citep{Oswalt1991}.

\subsubsection{DZ results}
\label{sec:dz_fits}

\begin{figure*}
    \centering
	\includegraphics[width=\textwidth]{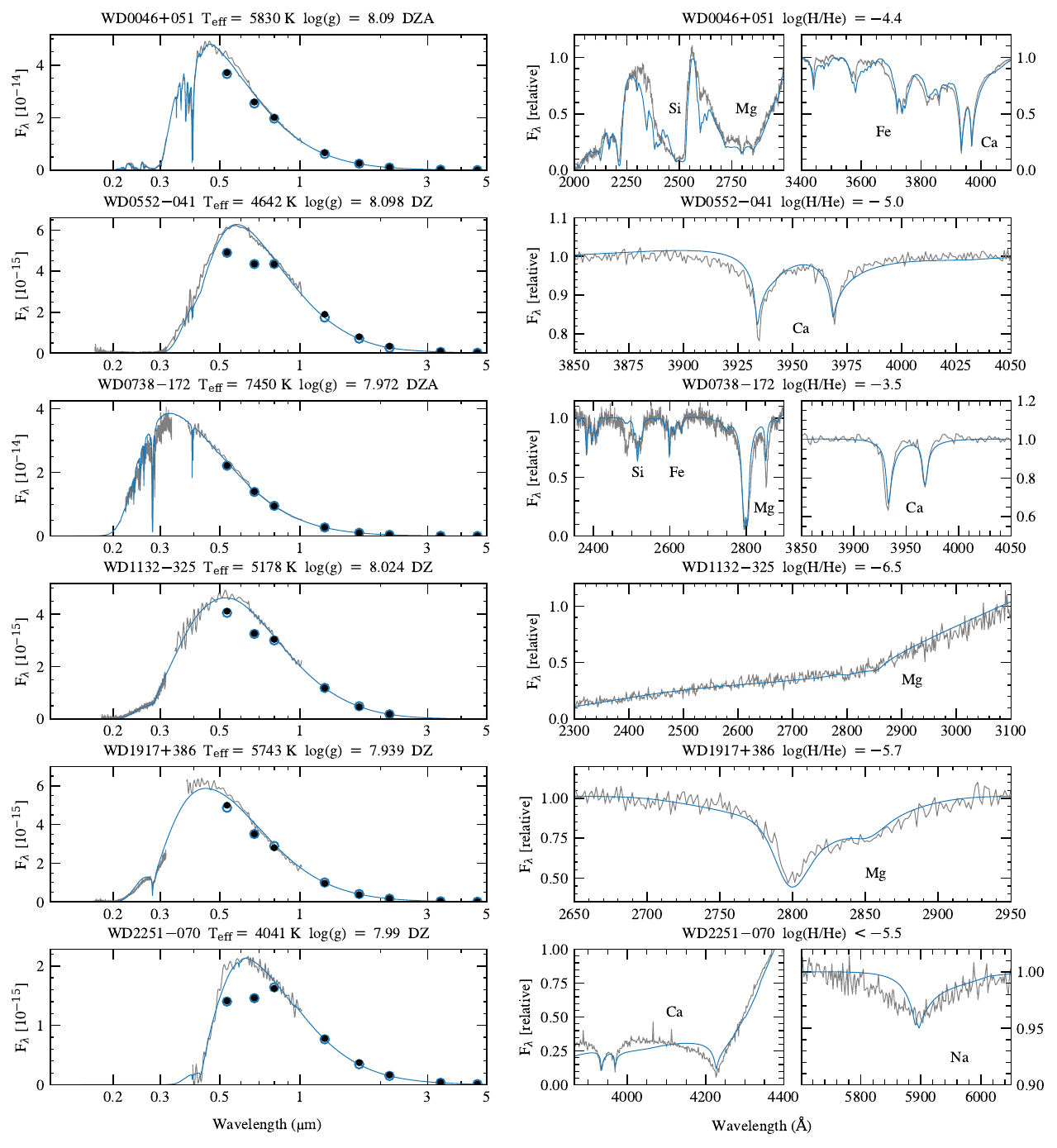}
	\caption{Combined spectrophotometric fits of DZ white dwarfs in the 13\,pc sample with metal absorption lines. \textit{Left}: combined fits to the full SED. \textit{Right}: fits to the metal features. Flux is given in units of erg\,cm$^{-2}$\,s$^{-1}$\,\AA$^{-1}$. The blue models show the overall fits, and the blue circles show how well these same models fit to the photometry (black dots).}
    \label{fig:plot_DZ}
\end{figure*}

The best-fitting parameters and abundances for the DZ white dwarfs within 13\,pc are presented in Table~\ref{tab:13pc_all_parameters_DZ}, and fits are shown in Fig.~\ref{fig:plot_DZ}. Each DZ white dwarf required tailored modelling, and we discuss them individually below.

\textbf{WD\,0046$+$051}: We classified WD\,0046$+$051 (van\,Maanen\,2) as a DZA owing to a very weak and narrow H\,$\alpha$ line in its X-shooter spectrum. The X-shooter spectrum also showed Ca\,\textsc{i}, Ca\,\textsc{ii}, Fe\,\textsc{i}, Cr\,\textsc{i}, Ni\,\textsc{i} and Na\,\textsc{i} features. The STIS spectrum displayed Fe\,\textsc{ii}, Si\,\textsc{i}, Mg\,\textsc{i} and Mg\,\textsc{ii} features. The \Teff\ and hydrogen content required for the observed weak and narrow H\,$\alpha$ feature were significantly discrepant with the \Teff\ and hydrogen abundance found from fitting the full SED and metal lines. Similar issues have been identified for H\,$\alpha$ in many cool, helium atmosphere metal enriched white dwarfs (see \citealt{OBrien2025} for details). Therefore, we opted not to fit H\,$\alpha$ in the spectrum of WD\,0046$+$051 and instead we constrained the hydrogen abundance from the combined SED fit and metal line widths. 

Using Lorentzian profiles for all silicon lines in the near-UV, we were unable to achieve an adequate fit to the STIS spectrum. The 2500\,\AA\ Si\,\textsc{i} multiplet showed a strong bluewards asymmetry suggesting broadening by neutral helium. In the absence of bespoke unified profiles, we used quasi-static profiles to model these lines, which provided an improved fit \citep{Walkup1982}. Our results motivate calculations for silicon lines broadened by neutral helium for future analyses. We believe this fit to the spectra and photometry of van Maanen 2 to be the best published fit to date.

\textbf{WD\,0552$-$041}: This white dwarf only showed Ca\,\textsc{ii}~H and K lines in the optical, and had no metal features in the UV. The STIS spectrum of this star had almost zero flux, due to its low \Teff\ and line blanketing from metals. An independent, detailed study of WD\,0552$-$041, using state-of-the art models to fit optical and UV spectra achieved a good fit to the Ca\,\textsc{ii}~H and K lines as well as the full SED, by incorporating trace magnesium in the models at a level that did not produce spectral features \citep{Blouin2018b}. Influenced by the \citet{Blouin2018b} fit, we also incorporated magnesium into our models at a level that did not produce features. We determined log(H/He)\,$=$\,$-$5, which provided a balance that reproduced the blue slope and limited the impact of CIA in the models.

\textbf{WD\,0738$-$172}: This is a well-known metal enriched white dwarf, which is warmer than the other DZ white dwarfs within 13\,pc, enabling a straightforward model fit. No STIS spectrum was available for this white dwarf, so we instead fitted a spectrum from \textit{HST} FOS \citep{Koester2000}. The optical spectrum showed Ca\,\textsc{ii}~H and K lines as well as an H\,$\alpha$ line, which we fitted alongside the SED to determine the hydrogen abundance. The UV spectrum displayed Fe\,\textsc{ii}, Si\,\textsc{i}, Mg\,\textsc{i} and Mg\,\textsc{ii} lines. The calcium and hydrogen abundances we determined are similar to recent measurements for this star \citep{Koester2000,Dufour2007,Coutu2019}. The feature that is not well-modelled at $\approx$\,2470\,\AA\ is likely from Fe\,\textsc{i}. 

\textbf{WD\,1132$-$325}: No metals were visible in the optical spectrum of this star, and therefore this white dwarf has historically been classified as a DC. However, in the STIS spectrum there was a very broad Mg\,\textsc{i} line. Following the \citet{Sion1983} spectral type classification system, this would be a `DC with UV metals'. However, this white dwarf is effectively part of the DZ class. The Mg\,\textsc{ii} doublet in the STIS spectrum was barely visible, and the spectrum is dominated by a broad Mg\,\textsc{i} `knee'. We measured the hydrogen abundance from the Ly\,$\alpha$ red wing and Mg\,\textsc{i} line width, and also determined an upper limit on the calcium abundance using an optical spectrum \citep{Giammichele2012}. 

\textbf{WD\,1917$+$386}: Similarly to WD\,1132$-$325, no metals were visible in the optical spectrum of this star, meaning up until now it had been classified as a DC. We re-classified it as a DZ, as it showed Mg\,\textsc{i} and Mg\,\textsc{ii} lines in the UV. We incorporated the STIS spectrum into the combined fit by masking the magnesium lines, and we measured the hydrogen abundance from the Ly\,$\alpha$ red wing. We also determined an upper limit on the calcium abundance by measuring the effect of calcium on the shape and strength of the magnesium lines.

\textbf{WD\,2251$-$070}: This DZ is so cool and UV-faint that the STIS G230L grating was unable to detect any near-UV flux. In the optical, there were Ca\,\textsc{i}, Ca\,\textsc{ii} and Na\,\textsc{i} lines. The optical spectrum also showed a broad unidentified feature near 4500\,\AA. A similar unexplained feature was identified in the cool DZ WD\,2356$-$209 \citep{Blouin2019}. Independent fits to the Ca\,\textsc{i} and Ca\,\textsc{ii} features of WD\,2251$-$070 constrained log(Ca/He)\,$=$\,$-$9.8 and log(H/He)\,$\leq$\,$-$4.5 \citep{Blouin2019, Blouin2019d}. Our best-fitting model in Fig.~\ref{fig:plot_DZ} required a substantial amount of magnesium for a good fit to the SED, similarly to WD\,0552$-$041. For Ca\,\textsc{i}, our unified profile did not reach deep enough, but did show the expected bluewards asymmetry. For sodium, the highly broadened line wings blended into the continuum, but a line core remained that was narrower than in the data, suggesting the upper atmosphere is even denser than in our already dense model. We found a limit of log(H/He)\,$<$\,$-$5.5, as above this abundance CIA features appear in the model that are not in the data. WD\,2251$-$070 is 4041\,K, making it one of the coolest known metal enriched white dwarfs \citep{Elms2022,OBrien2023,Kaiser2025}. The spectroscopic fit to WD\,2251$-$070 highlights the limitations of our models at low \Teff\ and high density.

\begin{table*}
    \centering
    \caption{Best-fit parameters for helium-atmosphere white dwarfs within 13\,pc of the Sun that are enriched by metals. Symbols indicate the data used for the fit. No symbol indicates STIS, \textit{Gaia}, 2MASS and WISE. $*$ indicates \textit{Gaia} and 2MASS. $\ddag$ indicates that no STIS spectrum was available for the combined fit.}
    \label{tab:13pc_all_parameters_DZ}
    \begin{tabular}{lllllll}
        \hline
        WD name & SpT & \Teff\ (K) & \logg & Mass (\Msun)& $\log$(H/He)& $\log$(Ca/He)\\
        \hline
        0046$+$051 & DZA & 5830\,$\pm$\,15 & 8.09\,$\pm$\,0.02 & 0.627\,$\pm$\,0.013& $-$4.4\,$\pm$\,0.1 & $-$10.14\,$\pm$\,0.03  \\
        0552$-$041 & DZ  & 4642\,$\pm$\,9 & 8.098\,$\pm$\,0.007 & 0.629\,$\pm$\,0.005 & $-$5.0\,$\pm$\,0.5 & $-$11.07\,$\pm$\,0.06  \\
        0738$-$172 $\ddag$ & DZA & 7450\,$\pm$\,87 & 7.972\,$\pm$\,0.029 & 0.559\,$\pm$\,0.017 & $-$3.50\,$\pm$\,0.05 & $-$10.83\,$\pm$\,0.02 \\
        1132$-$325 $*$ & DZ  & 5178\,$\pm$\,1 & 8.024\,$\pm$\,0.002 & 0.583\,$\pm$\,0.001 & $-$6.5\,$\pm$\,0.1 & $<-$12.5  \\
        1917$+$386 & DZ  & 5743\,$\pm$\,2 & 7.939\,$\pm$\,0.002 & 0.534\,$\pm$\,0.001 & $-$5.7\,$\pm$\,0.2 & $<-$13.5  \\
        2251$-$070 $\ddag$ & DZ  & 4041\,$\pm$\,17 & 7.99\,$\pm$\,0.02 & 0.561\,$\pm$\,0.012 & $<-$5.5 & $-$10.6\,$\pm$\,0.05  \\
        \hline
        WD name & $\log$(Mg/He)& $\log$(Fe/He)& $\log$(Si/He)& $\log$(Na/He)& $\log$(Cr/He)& $\log$(Ni/He)\\
        \hline
        0046$+$051 & $-$8.6\,$\pm$\,0.2 & $-$9.3\,$\pm$\,0.1 & $-$8.6\,$\pm$\,0.2 & $-$9.6\,$\pm$\,0.1 & $-$10.83\,$\pm$\,0.07 & $-$10.97\,$\pm$\,0.05 \\
        0552$-$041 & $-$9.5\,$\pm$\,0.5 & -- & -- & -- & -- & -- \\
        0738$-$172 $\ddag$ & $-$9.7\,$\pm$\,0.1  & $-$10.12\,$\pm$\,0.02 & $-$9.40\,$\pm$\,0.1 & -- & -- & -- \\
        1132$-$325 $*$ & $-$11.38\,$\pm$\,0.05 & -- & -- & -- & -- & -- \\
        1917$+$386 & $-$10.58\,$\pm$\,0.04& -- & -- & -- & -- & -- \\
        2251$-$070 $\ddag$ & $-$8.5\,$\pm$\,0.2 & -- & -- & $-$10.6\,$\pm$\,0.3 & -- & -- \\
        \hline
        
    \end{tabular}
\end{table*}

\subsection{Unknown composition white dwarfs}
\label{sec:magnetic_results}

\begin{figure*}
    \centering
	\includegraphics[width=\textwidth]{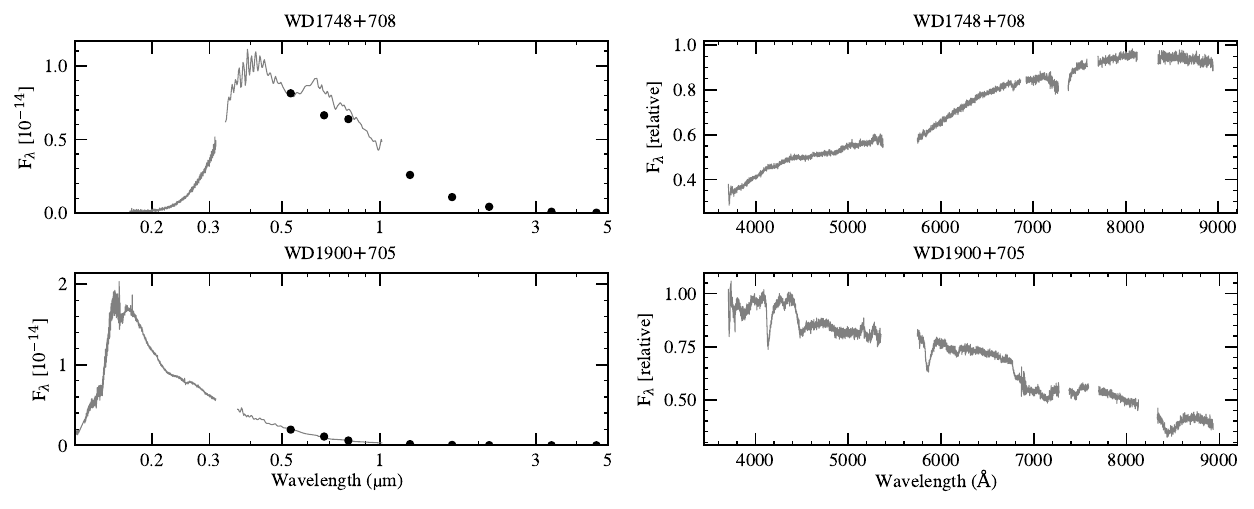}
	\caption{Spectra and photometry of strongly magnetic white dwarfs with unconstrained spectral types in the 13\,pc sample. \textit{Left}: STIS \gaia\ XP spectra, plus \gaia, 2MASS and WISE photometry. \textit{Right}: WHT ISIS optical spectra. Flux is given in units of erg\,cm$^{-2}$\,s$^{-1}$\,\AA$^{-1}$.}
    \label{fig:plot_MWDs}
\end{figure*}

Two of the white dwarfs in the 13\,pc sample, WD\,1748$+$708 and WD\,1900$+$705, are very strongly magnetic, with fields of 300\,MG and 200\,MG respectively \citep{Bagnulo2021}. The strong magnetic fields cause the spectral features of both white dwarfs to become distorted and blended, therefore making it challenging not only to fit the spectra but to identify the atmospheric composition of the stars. WD\,1900$+$705 (Grw$+$70$^{\circ}$8247) was the first white dwarf in which a magnetic field was detected \citep{Kemp1970}. A pure-H atmosphere for this star has been reported \citep{Ferrario2015}. The spectrum of WD\,1900$+$705 could not be fitted with state-of-the-art magnetic models \citep{Hardy2023,Hardy2023b}. WD\,1748$+$708 (G240$-$72) was also one of the first white dwarfs found to have a magnetic field \citep{Angel1974}, and the current consensus is that it has a helium-rich or carbon-rich atmosphere. Fig.~\ref{fig:plot_MWDs} shows the complex splitting in both white dwarf spectra. Both WD\,1748$+$708 and WD\,1900$+$705 display broad-band circular polarisation of their continua \citep{Berdyugin2022}. Due to their uncertain atmospheric compositions, we classified WD\,1748$+$708 and WD\,1900$+$705 as DXH in Table~\ref{tab:13pc_info}. Approximate parameters determined from non-magnetic model fits are: \Teff\,$=$\,5667\,K and \logg\,$=$\,8.35 for WD\,1748$+$708 and \Teff\,$=$\,11\,547\,K and \logg\,$=$\,8.53 for WD\,1900$+$705 \citep{Vincent2023}. New magnetic models could be used to fit these stars and determine their field strength and composition \citep{VeraRueda2024,Desai2025}.

\section{Discussion}
\label{sec:discussion}

\subsection{Combined fitting}
\label{sec:uv_vs_opticalir}

\citet{Saumon2014} analysed white dwarfs cooler than 6000\,K with STIS observations to test the effect of the Ly\,$\alpha$ red wing opacity in white dwarf models. They generated photometric datapoints by binning the STIS spectra, before fitting them in combination with optical and IR photometry. They compared optical plus IR photometric fits to STIS plus optical plus IR fits, and concluded that their parameters agreed within uncertainties and there was no systematic offset. They also found that incorporating STIS data into their fits halved the \Teff\ uncertainty. The parallaxes used by \citet{Saumon2014} were ten times less precise than the more recent \gaia\ parallaxes, and the analysis of \citet{Saumon2014} additionally differs from ours due to their binning of the STIS spectra. We opted to include the full information from the STIS spectra, since compared to \citet{Saumon2014} the 13\,pc sample provides a wider \Teff\ range and greater spectral diversity, including features such as carbon and Ly\,$\alpha$ lines. The two fits presented in our work $-$ STIS\,$+$\,\gaia\,$+$\,IR and \gaia\,$+$\,IR only $-$ are largely independent, as the former is primarily constrained by the UV spectra, except in the case of the coolest white dwarfs. We were therefore able to identify a discrepancy between the fits to UV data and to optical/IR data. The additional precision of \gaia\ photometry has also enabled reductions to the magnitudes of our \Teff\ and \logg\ fitting uncertainties compared to those presented in \citet{Saumon2014}. 

\begin{figure}
    \centering
	\includegraphics[width=\columnwidth]{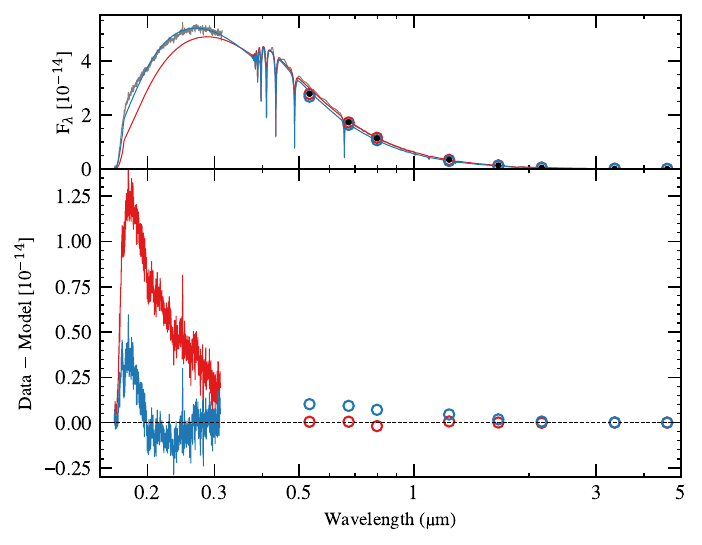}
	\caption{Fits and residuals for WD\,1202$-$232. Flux is given in units of erg\,cm$^{-2}$\,s$^{-1}$\,\AA$^{-1}$. The blue model is the best fit to STIS spectra plus \textit{Gaia}, 2MASS, and WISE photometry plus parallaxes. The red model is the fit to just the \textit{Gaia}, 2MASS, and WISE photometry plus parallaxes. The blue and red circles show how well these same models fit to the photometry. The upper panel shows the fits and the lower panel shows the residuals, comparing the models to the STIS spectra and photometry.}
    \label{fig:plot_residuals_wd1202}
\end{figure}

\begin{figure}
    \centering
	\includegraphics[width=\columnwidth]{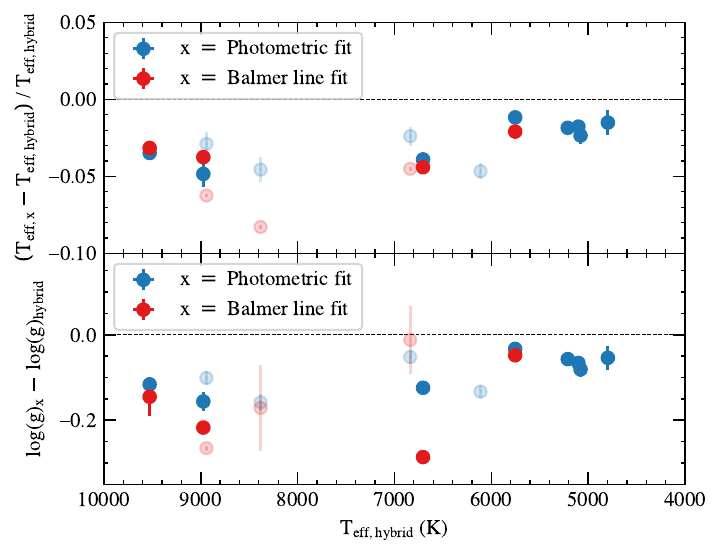}
	\caption{Comparisons of \Teff\ (top panel) and $\log g$ (bottom panel) for H-atmosphere white dwarfs fitted with different methods. The difference between the photometric optical and IR fit, compared to the parameters derived from the combined STIS G230L, optical photometry and IR photometry fit, is shown in blue. Also shown in red is the comparison between the Balmer line fit parameters and the combined parameters. Faded points correspond to magnetic white dwarfs.}
    \label{fig:plot_all_teff_logg_differences}
\end{figure}

Fig.~\ref{fig:plot_all_DA_G230L} demonstrates that neither our UV-dominated fits nor our optical plus IR fits are a good visual match across all wavelengths. A representative example is given in Fig.~\ref{fig:plot_residuals_wd1202}, which displays the residuals for fits to WD\,1202$-$232 with and without STIS data. This overall discrepancy is quantified across the sample in Fig.~\ref{fig:plot_all_teff_logg_differences} for both \Teff\ and \logg. For \Teff\,$<$\,10\,000\,K, which corresponds to the spectra with flux-calibrated STIS G230L data, the combined \Teff\ values are systematically 1\,--\,5\,per\,cent larger than the photometric \Teff\ values. The combined \logg\ values are also larger than the photometric \logg\ values by $\approx$\,0.07\,dex. These differences are smaller for the lowest \Teff\ members of the sample, because at these temperatures the STIS spectra have minimal flux and therefore do not provide much new information for the fit. All these fits relied on 3D model atmospheres (available for $\Teff\,>\,6000$\,K) that did not have a free parameter for the convective mixing-length. Through testing, we found that using standard 1D models instead of the 3D models did not solve the UV-optical discrepancy. A similar offset is seen in Fig.~\ref{fig:plot_all_teff_logg_differences} for the comparison of Balmer line fits to the combined fits \citep[see also][]{Sahu2023}. 

In contrast to the cooler white dwarf G230L observations, the hottest DA white dwarfs in the sample (Fig.~\ref{fig:plot_all_DA_echelle}) have substantially smaller combined \Teff\ values than photometric. An independent study found that for COS spectra of DA white dwarfs with temperatures of $12\,000<\Teff<33\,000$\,K, the COS \Teff\ values were 3\,per\,cent lower than optical Balmer line fits \citep{Sahu2023}. This result provides another hint that there may be a reversal in the trend of UV vs optical parameters at higher \Teff.

The so-called low-mass issue, which has been discussed extensively in previous studies, highlights that masses of white dwarfs cooler than 6000\,K were under-predicted by up to 30\,per\,cent when calculated using optical photometric fitting \citep{Hollands2018_Gaia,Caron2023,OBrien2024,Sahu2025}. Correcting the H$_3^+$ partition function in our models (see Section \ref{sec:models}) reduced the effects of the low-mass issue substantially, although a significant residual discrepancy remains at the lowest temperatures depending on the exact Ly\,$\alpha$ and CIA opacity sources used in the code. Modelling opacities in the UV-to-IR range for these cool white dwarfs remains a challenge \citep{Blouin2024,Sahu2025}, and hence issues with opacity modelling could still be a cause of the discrepancies between our fits with and without STIS data for \Teff\,$<$\,10\,000\,K.

The STIS low-dispersion gratings, including G230L which was used in our combined fits, were found to be flux calibrated to within 3\,per\,cent \citep{Elms2024}. \gaia\ photometry was also found to be typically uncertain by $\approx$\,0.01 mag \citep{Elms2024}. We therefore tested the impact of STIS flux calibration and \gaia\ photometric uncertainty on the resulting \Teff\ and \logg\ uncertainties determined from our combined fits to the 13\,pc DA white dwarfs. For this test we performed a Monte Carlo analysis, in which the STIS spectra error bars were inflated between their nominal values and  3\,per\,cent of the flux, and the \gaia\ magnitude error bars were inflated between their nominal values and 0.01 mag. These perturbations were both set to follow a normal distribution, and we performed 200 iterations of the least-squares fit with the perturbed error bars. The resulting 1\,$\sigma$ values were taken to be the \Teff\ and \logg\ uncertainties, independent of the statistical uncertainty determined initially from the least-squares fit. We found that introducing these additional observational systematic sources of uncertainty did not significantly increase the size of the error bars on the \Teff\ and \logg\ parameters compared to our quoted statistical error bars. We conclude that the model systematics caused by UV, optical, and IR opacities dominate in the error budget for white dwarfs in the \Teff\,$<$\,10\,000\,K regime.

\subsection{Carbon and hydrogen in He-atmosphere white dwarfs}
\label{sec:carbon_hydrogen}

According to spectral evolution theory, white dwarfs with helium-dominated atmospheres should have evolved from a common origin: a hot helium-rich DO white dwarf. The subsequent evolution of the spectral type, from DA/DB to DBA to DC/DQ, is mainly determined by the initial amount and distribution of hydrogen and carbon in the envelope (see \citealt{Bedard2024} for a review). In particular, the convective dredge-up of carbon is expected to occur in all helium-rich objects as they cool, but only those with the thinnest primordial helium envelopes become DQ white dwarfs. The others never display observable carbon features and thus become DC or DZ white dwarfs. This scenario predicts a continuous range of surface carbon abundances for the combined DC$-$DQ$-$DZ population at a given temperature. Convective dredge-up is also expected to be more efficient in lower-mass objects, which may explain why classical DQ white dwarfs tend to have slightly lower-than-average masses \citep{Bedard2022_iii}. However, hydrogen and metals can increase the opacity of the atmosphere and reduce the size of the convective zone, thus making it more challenging to detect carbon. Therefore, it could also be the case that DC and DZ white dwarfs have similar quantities of carbon to DQ white dwarfs but have more hydrogen and/or metals that suppress carbon features \citep{Blouin2022}.

The UV coverage of our sample enabled us to place an upper limit on the abundance of carbon in the DC white dwarfs (see Section~\ref{sec:dc_fits}). Fig.~\ref{fig:carbon_abundances_and_limits} shows the carbon abundances of DQ white dwarfs as a function of \Teff. The carbon limits for the DCs are shown in context of the DQ population in Fig.~\ref{fig:carbon_abundances_and_limits}. The carbon limits determined from the DCs were based on their UV spectra, whereas the DQ carbon abundances were determined from fitting only the optical Swan bands, so they are not directly comparable. Some DQs with purely UV carbon detections \citep{Weidemann1995} are indicated by hollow black circles on Fig.~\ref{fig:carbon_abundances_and_limits}. These points lie below the main DQ cluster along with our DCs. The DC carbon limits are also model-dependent, as at these lower \Teff\ values the optical limits from two different models that were plotted in Fig.~\ref{fig:carbon_abundances_and_limits} vary by $\approx$0.5\,dex. 

As discussed in Section~\ref{sec:dq_fits} we also determined the abundances or upper limits on the hydrogen content of the DQ white dwarfs. The measured hydrogen abundances are listed in Table~\ref{tab:13pc_all_parameters_DQ}, and for the remaining systems we found a consistent upper limit of $\log$(H/He)\,$=$\,$-$5.5. These limits are in line with the typical hydrogen abundances of DC white dwarfs (see Section~\ref{sec:dc_fits} and \citealt{Bergeron2019}).

To conclude, in the small sample of stars that we tested, we found that there was no significant difference between the hydrogen content of DQ and DC white dwarfs if we took the upper limits of hydrogen in DQs to be representative of their abundances. On the other hand, we found that the DC carbon limits were well below the DQ carbon abundances at the same \Teff\ values. The model issues identified in the UV for DQ white dwarfs, discussed in Section~\ref{sec:dq_fits}, make the direct comparison of carbon abundances derived here from UV (limits in DC) and optical spectroscopy (DQ) difficult. The DCs have an average mass of 0.69\,\Msun, while the DQs have an average mass of 0.54\,\Msun, so this may be a first hint at an explanation for their different natures. We note that this sample size is very small, and tests on larger datasets should be carried out.

\subsection{Radiative vs convective models}
\label{sec:magnetic}

All the white dwarfs in this work were fitted with convective atmosphere models, in regimes where convection is relevant \citep[see e.g., Figs.\,7 and 11 from][]{Saumon2022}. However, for a white dwarf harbouring a magnetic field with a strength above a few tens of kG to MG depending on the atmospheric composition, convective energy transfer is expected to be suppressed in the photosphere \citep{Tremblay2015_mag,Bedard2017,Gentile2018,Cunningham2021}. Therefore, a priori, radiative models should be adopted when fitting the observed spectral energy distributions of magnetic white dwarfs. We note that the effects of magnetic fields on macroscopic and atomic diffusion are much less important \citep{Cunningham2021}, and so this does not apply to the mixed masses, metal abundances or dredge-up of carbon. 

We tested the impact of using 1D radiative models for the combined fits of magnetic H-atmosphere white dwarfs compared to the convective models. The convective models are described in detail in Section~\ref{sec:Hatmosphere}. The radiative models are from \citet{Tremblay2015_mag} with line profiles from \citet{Tremblay2009}. We performed these tests on the four magnetic DA white dwarfs with STIS spectra in the 13\,pc sample: WD\,0009$+$501, WD\,0553$+$053, WD\,1953$-$011, and WD\,2359$-$434. The field strengths of these stars are 250\,kG, 15\,MG, 500\,kG, and 100\,kG respectively. For the coolest pure-H atmosphere white dwarfs at 4000\,K, convective energy transfer should be suppressed for field strengths above 30\,kG \citep{Cunningham2021}. At higher temperatures the minimum field strength for convection suppression decreases.

For a case study of WD\,2105$-$820 (a 10\,000\,K DAH) the radiative models provided a better fit \citep{Gentile2018}. However, WD\,1953$-$011, WD\,2359$-$434 and two other magnetic DAs not considered in this work were found to fit better in the optical with convective models \citep{Bedard2017}. Uniformly, the magnetic white dwarfs we tested had higher reduced $\chi^2$ values when fitted with the 1D radiative models, indicating a somewhat worse fit compared to the 3D convective models. The parameters determined from these fits have differences of up to 500\,K in \Teff\ and 0.15\,dex in \logg, which is consistent with the differences between parameters determined from optical and from UV fitting with the same models, as discussed in Section~\ref{sec:uv_vs_opticalir}. Therefore, despite the  reduced $\chi^2$ results, the convective models are not clearly preferred over the radiative models from our limited test set of magnetic white dwarfs. This subject requires a dedicated and detailed study with a larger set of magnetic DA white dwarfs.

\subsection{Planetary material}

Five white dwarfs in the 13\,pc sample show metal lines in their UV spectra but not in their optical spectra: WD\,0310$-$688 (DAZ), WD\,0435$-$088 (DQZ), WD\,1132$-$325 (DZ), WD\,1620$-$391 (DAZ), and WD\,1917$+$386 (DZ). These systems, spanning 5500\,K\,$-$\,25\,000\,K, highlight a major benefit of UV spectra. WD\,1132$-$325, and WD\,1917$+$386, which are featureless in the optical and show magnesium lines in their STIS spectra, have lower magnesium abundances than any other DZ white dwarf in the Planetary Enriched White Dwarf Database \citep{Williams2024}. The upper limits on the calcium abundances of these two stars firmly place them as some of the least polluted white dwarfs known \citep{Elms2022}. The low abundances of these cool DZs with UV-only metals highlight the importance of observing optical DC white dwarfs in the near-UV.

\subsubsection{Composition of accreted material}

We analysed the abundances of the 13\,pc metal enriched white dwarfs using \textsc{PyllutedWD} \citep{Harrison2018,Harrison2021,Buchan2022}, a Bayesian inference code that determines the geological histories of planetesimals that have been accreted by white dwarfs. \textsc{PyllutedWD} quantifies the evidence for key formation processes, including incomplete condensation, which refers to the depletion of volatile metals (such as oxygen) in planetary bodies due to the ambient temperature being high enough to prevent their condensation. The results may be sensitive to the metal sinking timescale grids and the inclusion of thermohaline mixing \citep{Buchan2025}. For the He-dominated WDs, we used the timescale grids of \citet{Koester2020} as these extend to 4000\,K. We excluded convective overshoot following the recommendations of \citet{Buchan2025}. For the H-dominated WDs, we used grids based on the \textsc{STELUM} evolution code in its static mode, including 3D calibrated overshoot \citep{Cunningham2019,Bedard2022_ii} for WDs at relevant temperatures (WD\,0310$-$688 in this case), or a parametrised treatment of overshoot otherwise, as in \citet{Buchan2025}. We included thermohaline mixing for systems warmer than 10\,000\,K. These assumptions made no qualitative impact unless otherwise noted. We added a noise floor of 0.1\,dex in quadrature to the metal abundance uncertainties quoted in Tables~\ref{tab:13pc_all_parameters_DAZ} and \ref{tab:13pc_all_parameters_DZ}, to avoid overfitting.

The planetesimal masses and accretion rates inferred by \textsc{PyllutedWD} are shown in Table~\ref{table:pyllutedwd}. The accretion rates in Table~\ref{table:pyllutedwd} correspond to the mean accretion rate over the presumed accretion episode, as opposed to the diffusion flux out of the convection zone. These two quantities are often assumed to be equal for DAZ white dwarfs, but this is not the case for DZs. The metal enriched white dwarfs that only have a single metal detection or upper limit (WD\,0141$-$675 and WD\,1202$-$232) were not analysed further, because a single metal abundance cannot constrain any geological processes. The metal enriched white dwarfs that have two or more metal detections and/or upper limits are discussed below. In all cases, the photospheric metal abundances could be accounted for with familiar stellar and geological processes, with the possible exception of extreme photospheric carbon depletion at WD\,1620$-$391.

\begin{table}
    \centering
        \caption{Masses and accretion rate outputs from the \textsc{PyllutedWD} framework for metal enriched white dwarfs within 13\,pc. Timescale grid abbreviations: \textit{S3D}\,$=$\,\textsc{STELUM} evolution code in its static mode, including 3D calibrated overshoot \citep{Bedard2022_ii}, with \textit{+T} indicating that thermohaline mixing was also included, \textit{KN}\,$=$\,Koester envelope code with no overshoot \citep{Koester2020}.}
    \begin{tabular}{llll}
    \hline
         WD Name&  log(Mass/g)&  log(Accretion rate/gs$^{-1}$)& Timescale\\
 & & &grid\\
         \hline
         0046$+$051&  21.4$^{+0.2}_{-0.2}$&  11.2$^{+1.8}_{-1.8}$& \textit{KN}\\
         0141$-$675 &  18.5$^{+0.4}_{-0.2}$&  9.0$^{+1.2}_{-1.6}$& \textit{S3D}\\
         0310$-$688&  16.8$^{+0.4}_{-0.2}$&  7.8$^{+0.1}_{-0.1}$& \textit{S3D+T}\\
         0552$-$041 &  19.8$^{+0.5}_{-0.5}$&  9.2$^{+2.0}_{-2.2}$& \textit{KN}\\
         0736$+$053&  19.0$^{+0.4}_{-0.3}$& 8.2$^{+2.2}_{-2.4}$& \textit{KN}\\
         0738$-$172&  20.3$^{+0.3}_{-0.2}$&  9.8$^{+2.1}_{-2.3}$& \textit{KN}\\
         1132$-$325 &  19.0$^{+0.5}_{-0.3}$& 8.0$^{+2.3}_{-2.6}$& \textit{KN}\\
         1202$-$232 &  17.4$^{+0.5}_{-0.6}$&  8.2$^{+1.2}_{-1.3}$& \textit{S3D}\\
         1620$-$391 &  16.9$^{+0.5}_{-0.3}$&  8.9$^{+0.2}_{-0.2}$& \textit{S3D+T}\\
         1917$+$386 &  22.0$^{+0.3}_{-0.3}$&  11.6$^{+1.7}_{-2.2}$& \textit{KN}\\
         2251$-$070&  21.1$^{+0.4}_{-0.5}$&  10.4$^{+1.9}_{-2.1}$& \textit{KN}\\
         \hline
    \end{tabular}
    \label{table:pyllutedwd}
\end{table}

Two DAZ white dwarfs were analysed with \textsc{PyllutedWD}. WD\,0310$-$688 is most likely to be accreting in the steady state phase, meaning that accretion is ongoing. The low oxygen abundance in the atmosphere of WD\,0310$-$688 implies incomplete condensation to a nominal confidence level of 2.2\,$\sigma$, but whether this process is invoked depends on the sinking timescale grid used and whether thermohaline mixing is included. The material accreted by  WD\,1620$-$391 could not be modelled accurately by \textsc{PyllutedWD}, because the carbon is heavily depleted relative to silicon, after normalising to the Solar composition. This depletion does, however, imply incomplete condensation. 

Six DZ(A) white dwarfs in 13\,pc were also analysed with \textsc{PyllutedWD}. In each case, \textsc{PyllutedWD} was able to fit the abundances well, and found that the systems had likely reached the declining phase of accretion (with at least 93\,per\,cent confidence). The low Ca/Mg ratio for WD\,1917$+$386 implies that it is particularly deep into the declining phase: $39\pm3$\,Myr since the end of accretion, or $7.0\pm0.5$\,$\tau_{\rm Mg}$, where $\tau_{\rm Mg}$ is the Mg sinking timescale. Given that accretion is inferred to have ceased, the total mass accreted by each white dwarf could be constrained. These masses range from $10^{19}$ to $10^{22}$ g, corresponding to asteroids or small moons. The accretion rates are relatively poorly constrained as they are sensitive to the assumed duration of accretion events.

We additionally analysed the DQZ white dwarf WD\,0736$+$053 (Procyon\,B) with \textsc{PyllutedWD}. The abundances are close to chondritic, and the most likely phase of accretion is the declining phase, with a probability at least 85\,per\,cent.

\subsubsection{Metal enrichment demographics}

Statistics of white dwarf samples accreting planetary debris depend on the sample selection effects (e.g. average \Teff\ and mass, hence characteristic age), atmospheric composition, spectroscopic resolution, and wavelength coverage. A sample of high-resolution optical spectra centred around the Ca\,\textsc{ii}~H and K lines revealed a fraction of 25\,per\,cent with metal lines \citep{Zuckerman2003}. A warm (17\,000\,K $<$ \Teff\ $<$ 27\,000\,K) white dwarf sample with far-UV spectroscopy showed that 27\,--\,50\,per\,cent of white dwarfs were metal enriched \citep{Koester2014}. An updated far-UV study revealed that at least 40\,per\,cent of warm (13\,000\,K $<$ \Teff\ $<$ 30\,000\,K) white dwarfs are metal enriched \citep{OuldRouis2024}. It has also been observed that the enrichment fraction decreases significantly for larger white dwarf masses \citep{Mccleery2020,OuldRouis2024,Cunningham2024}. For the 13\,pc sample, we found an enrichment fraction of $k/N$\,$=$\,13/44\,$=$\,30\,$\pm$\,7\,per\,cent. The 13\,pc sample has a median \Teff\ of 6700\,K, much lower than those of the aforementioned samples \citep{Zuckerman2003,Koester2014,OuldRouis2024}. Our marginally reduced fraction compared to \citet{OuldRouis2024} may be caused by the effects of metal line visibility in our sample, given the higher average mass of a volume-limited sample compared to a magnitude-limited sample \citep{Tremblay2016}, but could also hint at a decrease of enrichment rates as a function of cooling age, as the 13\,pc sample contains on average older white dwarfs \citep{Hollands2018_dz}.

The limiting sensitivity to accretion rates is broadly comparable between silicon detections in the UV (COS) and calcium detections in the optical (see Figure~8 of \citet{Koester2014}). If planetary system properties (e.g. the occurrence of planetary material, disc lifetimes, and the duty cycle of disruption/accretion events) are independent of age, then the enrichment incidence measured in the UV \citep[45$\,\pm\,$6\,per\,cent from][]{OuldRouis2024} should constitute a lower limit on the corresponding enrichment fraction in the 13 pc sample. This is because metal diffusion timescales are orders of magnitude longer in the (cooler) 13 pc sample, approaching ${\sim}10^6$ years, compared to timescales of order days in the COS sample \citep{Koester2020}. One thus expects to recover a similar fraction of systems in an active accretion phase, plus an additional population of white dwarfs observed during the declining phase. 

The enrichment fractions measured for the two samples are consistent at the 1.5\,$\sigma$ level. This is compatible with age-independent enrichment properties, but also with scenarios in which the frequency of accretion events (or the time-averaged accretion rate) decreases with time. In that case, a reduction in accreted mass would tend to lower the polluted fraction, while the increase of diffusion timescales with age would act in the opposite sense by extending the detectable ``memory'' of past accretion. The net effect is therefore degenerate: although it could, in principle, be explored with forward modelling, the size of the declining-phase population would depend sensitively on poorly constrained parameters (e.g. disk lifetimes and the interval between accretion episodes), making it difficult to disentangle with current data.

\subsection{Stellar multiplicity}

Table~\ref{tab:13pc_binaries} shows all confirmed 13\,pc multiple-star systems containing white dwarfs. The UV has historically been significant for detecting white dwarf companions to main sequence stars, since white dwarfs are typically brighter at UV wavelengths, relative to optical wavelengths, than main sequence stars. These multiple-star systems have been identified using proper motion techniques and radial velocity measurements. We find 14 multiple-star systems out of a total of 42 systems, where each WD\,$+$\,WD pair is counted as a single system. Therefore the white dwarf multiplicity fraction in 13\,pc is $k/N$\,$=$\,14\,/\,42\,$=$\,33\,$\pm$\,7\,per\,cent. The frequency and diversity of these systems indicate that they must be common in the Solar neighbourhood, given that one in three white dwarfs are part of multiple star systems. Additionally, some of the white dwarfs within 13\,pc are highly massive and/or strongly magnetic, with fields over 100\,MG, and such systems could be the products of stellar mergers \citep{Temmink2020,Bagnulo2022}. 

Using analysis from the 40\,pc white dwarf sample \citep{OBrien2024}, we determined a multiplicity fraction in the 13\,--\,40\,pc shell of 19.5\,per\,cent. We used this shell to determine an independent multiplicity fraction that does not contain the 13\,pc white dwarfs. Twenty per\,cent of the 13\,pc white dwarf sample consists of 9 stars, compared to the 14 multiple-star systems with a white dwarf within 13\,pc. Double white dwarf systems are detectable to larger distances in the optical due to their double-lined spectra or systematically low \logg\ values when fitted with single-star models. However, close white dwarf $+$ main sequence binaries are more challenging to detect in the optical due to the main sequence component outshining a cool white dwarf companion. The proximity of the white dwarfs in the 13\,pc sample makes the binary systems more easily resolvable, and the space-based near-UV spectra available for the sample also provide a dataset uncontaminated by companions. 

Without a comprehensive radial velocity survey of the whole sky, many of our systems, if placed at a distance of 40\,pc, would be missed with only optical observations. The white dwarfs in the Sirius, Procyon, 40\,Eri, GJ\,86, and G\,203-47 systems were more easily identified due to the proximity of these systems to the Sun, enabling the white dwarf components to be resolved by \textit{HST}. If these systems were further than a few pc away, the white dwarfs would be challenging to identify, let alone observe and characterise, due to the brightness of their main sequence companions. Additionally, if they were placed at $\approx$\,100\,pc, Sirius, Procyon, 40\,Eri, and GJ\,86 would not be resolved by \gaia, which has an on-sky resolution of 0.4\,arcsec. 

\begin{table*}
    \centering
        \caption{Details of all confirmed white dwarfs within 13\,pc of the Sun that are part of a multiple-star system. Projected separations were calculated in most cases using \gaia\ DR3 data, aside from those binaries that are unresolved in \gaia. A colon symbol indicates that the spectral type of the star has not been confirmed. For higher-order systems (triple or quadruple) the periods separated by semicolons are the innermost to outermost periods for the hierarchical system.}
    \begin{tabular}{lllllllll}
    \hline
         Star 1&  Star 1&  Star 2&  Star 2&  Star 3&  Star 3 & Star 4&Star 4& Projected\\
         Name&  SpT&  Name&  SpT&  Name&  SpT & Name&SpT & Separation [au]\\
         \hline
         Sirius A&  A0&  WD\,0642$-$166&  DA&  --&  -- & --&-- & 20\\
         Procyon A&  F5&  WD\,0736$+$053&  DQZ&  --&  -- & --&-- & 15\\
         40 Eri A&  K1&  WD\,0413$-$077&  DA&  40 Eri C&  M4.5 & --&-- & 42; 417\\
         Stein 2051 A&  M4&  WD\,0426$+$588&  DC&  --&  -- & --&-- & 57\\
         G 203-47 A& M3.5 & G 203-47 B& DC & --& -- & --&-- &0.05\\
         IRAS 21500$+$5903& M3& WD\,2150$+$591& DAH& --& -- & --&-- &124\\
         LP 783-2& M6.5& WD\,0738$-$172& DZA& --& -- & --&-- &186\\
         GJ 432 A& K0& WD\,1132$-$325& DZ& --& -- & --&-- &146\\
         L 923-22& M5& WD\,1917$-$077& DBQA& --& -- & --&-- &285\\
         GJ 86 A& K1.5& WD\,0208$-$510& DQ& --& -- & --&-- &28\\
         WD\,0727$+$482 A& DA& WD\,0727$+$482 B& DA& G 107-69 A&  M4.5& G 107-69 B&dM/BD:&0.5; 8; 1161\\
         GJ 1179 A& M5& WD\,1345$+$238& DC& --&  -- & --&-- &2231\\
         WD\,0135$-$052 A& DA&  WD\,0135$-$052 B& DA& --& -- & --&-- &0.03\\
         HD 147513& G5& WD\,1620$-$391& DA& --&  -- & --&-- &4454\\
 \hline
    \end{tabular}
    \label{tab:13pc_binaries}
\end{table*}

\subsection{Variability}

White dwarfs show photometric variability for a variety of reasons, including magnetism, pulsations, and binarity. We checked all 13\,pc white dwarfs to see if they showed variability in \textit{Transiting Exoplanet Survey Satellite} (\textit{TESS}; \citealt{Ricker2015}) lightcurves. All signals and non-variability limits for the 13\,pc white dwarfs are listed in Table~\ref{tab:variability}. Most of the sample do not show variability, to very good limits. Four of the magnetic white dwarfs show significant \textit{TESS} periods, most likely due to spots at the rotation period. They are: WD\,2359$-$434 at 2.695\,hours, WD\,0009$+$501 at 8.02\,hours, WD\,0912$+$536 at 31.95\,hours, and WD\,1953$-$011 at 34.62\,hours. The rotation periods of the first three of these magnetic systems were identified via a crowding analysis to check that the signal originated from the white dwarf \citep{Oliveira2024}, using \textsc{TESS-localize}\footnote{\url{https://github.com/Higgins00/TESS-Localize}} \citep{Higgins2023}. The results from \textsc{TESS-localize} are ambiguous for WD\,1953$-$011, but it was flagged as photometrically variable in \gaia\, which shows the \textit{TESS} period, therefore confirming that the signal is coming from the white dwarf \citep{Steen2024}. Additionally, the well-known pulsator WD\,1647$+$591 has a rotation period of 9.1\,hours determined from frequency splittings in the \textit{TESS} data \citep{Bognar2024}. A few other white dwarfs appeared to show significant \textit{TESS} variability signals, WD\,1620$-$391 at 3.44\,hours and WD\,0839$-$327 at 11.43\,hours, but after careful exploration with \textsc{TESS-localize}, the signals did not appear to be coming from the white dwarf but rather another nearby star within the \textit{TESS} field of view. 

\begin{table}
    \centering
        \caption{Periods or variability limits from \textit{TESS} data of 13\,pc white dwarfs. \textsc{TESS-localize} was applied to check if any signal originated from the white dwarf. An NOV of 1 means the source is not variable to 1\,ppt (part-per-thousand), equivalent to 0.1\,per\,cent amplitude. Magnetic field strengths are also listed \citep{Bagnulo2021}}.
    \begin{tabular}{llll}
         \hline
         WD &$P_{\rm rot}$ (hr) &NOV (ppt) &Magnetic field\\
         name & & &strength (MG)\\
        \hline
         0009$+$501 & 8.02&[2.5 ppt amp. spot] &0.25\\
         0038$-$226 & --&1 &--\\
         0046$+$051 & --&0.2 &--\\
         0135$-$052 & --&0.2 &--\\
         0141$-$675 & --&0.2 &--\\
         0208$-$510& --&Field too crowded &--\\
         0245$+$541 & --&1 &--\\
         0310$-$688 & --&0.1 &--\\
         0413$-$077 & --&Field too crowded &--\\
         0426$+$588& --&Field too crowded &--\\
         0435$-$088 & --&0.5 &--\\
         0548$-$001 & --&0.5 &7\\
         0552$-$041 & --&0.5 &--\\
         0553$+$053 & --&1.5 &15\\
         0642$-$166& --&Field too crowded &--\\
         0727$+$482\,A& --&0.3 &--\\
         0727$+$482\,B& --&0.3 &--\\
         0736$+$053& --&Field too crowded &--\\
         0738$-$172& --&0.3 &--\\
         0752$-$676 & --&0.1 &--\\
         0810$-$353& --&1 &30\\
         0821$-$669 & --&0.4 &--\\
         0839$-$327 & --&Signal not from WD &--\\
         0912$+$536 & 31.95&[19.2 ppt amp. spot] &100\\
         1055$-$072 & --&No \textit{TESS} data available &--\\
         1132$-$325 & --&Field too crowded &--\\
         1142$-$645 & --&Signal not from WD &--\\
         1202$-$232 & --&0.2 &--\\
         1334$+$039 & --&0.7 &--\\
         1345$+$238 & --&1.5 &--\\
         1620$-$391 & --&Signal not from WD &--\\
         1630$+$089& --&No \textit{TESS} data available &--\\
         1647$+$591 & 9.1&[1.8 ppt amp. pulsations] &--\\
         1748$+$708 & --&0.1 &300\\
         1900$+$705 & --&0.1 &200\\
         1917$-$077 & --&1 &--\\
         1917$+$386 & --&0.6 &--\\
         1953$-$011 & 34.62&[25.0 ppt amp. spot] &0.5\\
         2140$+$207 & --&0.4 &--\\
         2150$+$591 & --&Field too crowded &0.8\\
         2251$-$070 & --&1.5 &--\\
         2359$-$434 & 2.695&[3.7 ppt amp. spot] &0.1\\
         \hline
    \end{tabular}
    \label{tab:variability}
\end{table}

\section{Conclusions}

In this work, we analysed near-UV, optical, and IR observations of white dwarfs within 13\,pc of the Sun. The main observations underpinning this work were from the STIS instrument onboard \textit{HST}. The volume-limited 13\,pc sample consists primarily of cool white dwarfs, and the STIS observations provided flux-calibrated UV data for testing and calibrating atmosphere models. The STIS spectra enabled us to identify many new or unexpected features in the atmospheres of the white dwarfs in the sample, including the broad Mg\,\textsc{ii}\,2800\,\AA\ lines identified in He-rich white dwarfs with low-resolution spectra, and narrow metal lines identified in H-rich warm white dwarfs. Specifically, we identified a Mg\,\textsc{ii} feature in the STIS spectra of WD\,1917$+$386 (previously DC) and WD\,0435$-$088 (previously DQ), and a Mg\,\textsc{i} feature in the STIS spectrum of WD\,1132$-$325 (previously DC). The following white dwarfs in the sample only display metal features in the UV: WD\,0310$-$688 (DAZ), WD\,0435$-$088 (DQZ), WD\,1132$-$325 (DZ), WD\,1620$-$391 (DAZ), WD\,1917$+$386 (DZ). Therefore optical spectroscopy alone would miss five metal enriched white dwarfs in the sample. Additionally, carbon features are only seen in the UV for the DBQA white dwarf WD\,1917$-$077, and H$_2$ molecular lines are detectable only in the UV for the pulsating DAV WD\,1647$+$591.

Following our fits to the 13\,pc spectra and photometry, we compared UV and optical parameters for H-atmosphere white dwarfs. We found that for $\Teff < 10\,000$\,K the \Teff\ from UV fitting was between 1 and 5\,per\,cent larger than the \Teff\ from optical fitting, showing that current models fail to replicate the full SEDs of these white dwarfs. The flux-calibrated STIS spectra of the He-rich DC white dwarfs enabled us to constrain their hydrogen content, which is not straightforward with optical photometry alone. We identified a significant discrepancy between models and observations of carbon lines in the near-UV, despite accurate modelling of the optical Swan bands. After altering the $gf$ values for the UV C\,\textsc{i} lines to match the observations, we found systematically low masses from fitting UV observations of DQs, that are not consistent with the more typical masses of the same DQs determined from optical fitting. We analysed the difference between the hydrogen content and carbon content of DC and DQ white dwarfs. There was no measurable difference between the hydrogen content of DQs and DCs, but the upper limits of carbon in DCs were well below that of DQs. These results are model-dependent, and given the issues with modelling carbon lines in the UV, are treated with caution. The atomic data of the UV carbon lines should be revisited in order to improve future modelling efforts. We also tested the effect of fitting magnetic white dwarf spectra with radiative models as opposed to convective models, which should be a more accurate treatment of a magnetic white dwarf atmosphere. We found that neither the radiative nor convective models provided a significantly better fit. Finally, we found a 33\,per\,cent multiplicity fraction for the 13\,pc white dwarfs, and at least 30\,per\,cent of the white dwarfs showed evidence of accretion of planetary debris. 

The sample reveals potential avenues to improve white dwarf modelling across the UV-to-IR SED. The strength of carbon bands in the UV in current models is substantially different from observations. Metal lines in cool He-rich atmosphere white dwarfs are challenging to model, both in the UV and in the optical, and improvements must be made to the line profiles and other physics such as pressure ionisation. Observations of more optically classified DC white dwarfs would be beneficial, to reveal Mg\,\textsc{i} or Mg\,\textsc{ii} features, or possible carbon lines. Flux-calibrated \textit{JWST} NIRSpec and MIRI spectra for 35 white dwarfs within 13\,pc  will be collected as part of Cycle 5 (program 10131), which will provide insight into the nature of CIA in cool white dwarfs, as well as test the optical-UV \Teff\ discrepancy that has been identified from spectral modelling. 

\section*{Acknowledgements}

We thank the referee for their helpful comments and feedback. MOB, PET, BTG, SS and AMB received funding from the European Research Council under the European Union’s Horizon 2020 research and innovation programme numbers 101002408 (MOB, PET and AMB), and 101020057 (BTG and SS). MAH and AB acknowledge Warwick Astrophysics prize post-doctoral fellowships made
possible thanks to a generous philanthropic donation.

Support for the NASA/ESA Hubble Space Telescope Program number 14076 was provided by NASA through a grant from the Space Telescope Science Institute, which is operated by the Association of Universities for Research in Astronomy, Incorporated, under NASA contract NAS5-26555. This work has made use of data from the European Space Agency (ESA) mission \textit{Gaia} (\url{https://www.cosmos.esa.int/gaia}), processed by the \textit{Gaia} Data Processing and Analysis Consortium (DPAC, \url{https://www.cosmos.esa.int/web/gaia/dpac/consortium}). Funding for the DPAC has been provided by national institutions, in particular the institutions participating in the \textit{Gaia} Multilateral Agreement. 

This research has made use of the SVO Filter Profile Service "Carlos Rodrigo", funded by MCIN/AEI/10.13039/501100011033/ through grant PID2020-112949GB-I00. This research has made use of NASA’s Astrophysics Data System; the SIMBAD database, operated at CDS, Strasbourg, France; and the VizieR service. This work made use of \textsc{astropy}:\footnote{\url{http://www.astropy.org}} a community-developed core \textsc{python} package and an ecosystem of tools and resources for astronomy \citep{astropy:2013, astropy:2018, astropy:2022}. This work also made use of the \textsc{python} package \textsc{scipy} \citep{2020SciPy-NMeth}. Computing facilities were provided by the Scientific Computing Research Technology Platform of the University of Warwick. We thank S.~Blouin for sharing findings of the reason for the overestimation of the $\rm H_3^+$ partition function.

\section*{Data Availability}

\textit{HST} data are available in the MAST archive, with the majority of observations falling under Program ID 14076. ESO data are available in the ESO science archive, with the majority of observations falling under Program IDs 098.D$-$0392(A) and 099.D$-$0661(A). All observation details are in Table~\ref{tab:13pc_observations}. Some optical spectra used in this work were downloaded from the Montr\'eal White Dwarf Database \citep{Dufour2017}. Reduced WHT ISIS data are available on request.


\bibliographystyle{mnras}
\bibliography{mybib} 







\bsp	
\label{lastpage}
\end{document}